\documentclass[aps,prf,preprint,amsmath,amssymb,amsfonts]{revtex4-2}
\usepackage{dcolumn}
\usepackage{bm}
\usepackage{amsmath}
\usepackage{graphicx} %eps figures can be used instead`
\usepackage{epstopdf}%This line makes .eps figures into .pdf - please comment out if not required.
\usepackage{makecell}
\usepackage{hhline}

\usepackage{xcolor}
\newcommand{\be}{\begin{equation}}
\newcommand{\ee}{\end{equation}}
\newcommand{\bea}{\begin{eqnarray}}
\newcommand{\eea}{\end{eqnarray}}
\newcommand{\rIm}{\mathrm{Im}}
\newcommand{\rRe}{\mathrm{Re}}
\newcommand{\re}{\mathrm{e}}            % Roman e for exponential
\newcommand{\ri}{\mathrm{i}}
\usepackage[titletoc,title]{appendix}
\usepackage[normalem]{ulem} %To strikeout text

\begin{document}

\title{Quartz Crystal Microbalance impedance of discrete adsorbates with lubricated contact}

\author{Alexander M. Leshansky$^1$}\email{lisha@technion.ac.il} \author{Boris Y. Rubinstein$^2$} \author{Marta Sadowska$^3$}
\author{Zbigniew Adamczyk$^3$}

\affiliation{$^1\!$Wolfson Department of Chemical Engineering, Technion, Haifa 32000, Israel}
\affiliation{$^2\!$Stowers Institute for Medical Research, 1000 E 50th st., Kansas City, MO 64110, USA}
\affiliation{$^3\!$Jerzy Haber Institute of Catalysis and Surface Chemistry, Polish Academy of Sciences, 30-239 Krakow, Poland}

\begin{abstract}

We present theoretical investigation of the shear force exerted on a horizontal plate undergoing fast (MHz range) longitudinal oscillations in presence of a spherical particle suspended in the liquid at close proximity to the plate. This hydrodynamic problem is relevant toward the analysis of frequency  response (i.e., complex impedance) of Quartz Crystal Microbalance (QCM) to discrete (particulate) adsorbates. For low areal density of adsorbates, the contributions of individual particles remain additive, rendering the theory applicable towards quantitative interpretation of the experimental measurements. For adsorbates much smaller than the viscous penetration depth, the closed-form analytical expressions for the complex impedance were derived for particles of arbitrary density, assuming a lubricated contact (i.e., non-localized adsorption). The theoretical predictions were shown to be in an excellent agreement with accurate Finite Element numerical computations. The validity of the theoretical approach was confirmed by QCM measurements conducted for polystyrene nanoparticles in a broad range of sizes, from $32$~nm to $510$~nm, adsorbing onto gold/silica sensor upon varying the oscillation frequency.  

\end{abstract}

\maketitle

\section{Introduction}
Quartz crystal microbalance (QCM) technique relies on sharp piezoelectric resonance of AT-cut quartz crystal under application of a voltage \cite{book84,review21,review24}. The oscillations are in the form of an elastic shear wave with $1$--$2$~nm amplitude and MHz-range frequency. For example, the widely used commercial QCM device (QSense Analyzer, Biolin Scientific, Sweden) uses quartz crystal of thickness $334$~$\mu$m with the fundamental resonant frequency $f_0\!\simeq\!5$~MHz with overtones $f\!=\!n f_0$, where $n$ is an odd overtone number. The resonant frequency of the oscillations shifts to a lower value due to the mass of the adsorbed material, simply due to the increase of the inertial mass of the resonator, allowing extremely sensitive gravimetric measurements. In vacuum, the shift in the resonant frequency of the crystal is linearly proportional to the mass of the adsorbed film via the seminal Sauerbrey equation \cite{sauerbrey59}. Recall that QCM does not measure the adsorbed mass directly, but instead it measures the \emph{inertial force} due to the adsorbed mass; the Sauerbrey equation formally applies to measurements of rigid adsorbed films (i.e., fully coupled to the oscillations of the quartz sensor) in vacuum or in gas phase. In liquid-phase QCM, however, the fluid not only contributes to the shift in resonant frequency due to inertia of the fluid driven by the resonator, but it also affects the bandwidth of the signal, owing to viscous dissipation \cite{nomura80, qcm-liquid85}. The interpretation of the QCM-D measurement  (where ``D" stands for dissipation monitoring) in liquids is well-established for planar (including viscoelastic) adsorbed planar films \cite{film1,film2,film3,film4}. However, the quantitative interpretation of the QCM-D measurements involving \emph{discrete} adsorbates (e.g., liposomes, viruses, macromolecules, nanoparticles, bacteria, etc.) in liquids is still in progress, mostly due to complicated unsteady hydrodynamics, complex flow geometry and uncertainty associated with contact dynamics, which depends on physicochemical properties of the surfaces (i.e., the adsorbate and the sensor) and a solvent \cite{review21}. The experimental evidence showing considerable overestimation of the adsorbed mass based on the prediction of the Sauerbrey equation for discrete adsorbates has been known for quite a while, and the ``trapped liquid" \emph{ad hoc} phenomenological models were proposed to explain the apparent discrepancy (see, e.g., \cite{phenom1, phenom2, phenom3}).

The modern commercial QCM-D devices probe the area-averaged tangential stress, normalized by the velocity of the crystal oscillations, $\overline{\sigma}/v_c$, which is also known as impedance, ${\mathcal Z}$. The area-averaged tangential stress $\overline{\sigma}$ is determined as net shear force $\mathcal{F}_s$ exerted on the surface of the oscillating quartz sensor divided by its surface area $A$. Since in the frequency domain (i.e., Fourier space), both $\mathcal{F}_s$ and $v_c$ are complex quantities characterized by amplitude and phase, therefore the impedance ${\mathcal Z}$ is also a complex quantity. 
Typically in experiments the frequency shift is much smaller than the resonant frequency itself, $|\Delta f| \ll f_0$, and the ``small-load approximation" applies \cite{review21}, such that the shift in resonant frequency, $\Delta f$, and the shift in the half-bandwidth, $\Delta \Gamma$ (or the ``dissipation factor" $\Delta {\mathcal D}%\!=\!2\Delta \Gamma/f
$), are linearly proportional to the impedance,  $\Delta f-\ri \Delta\Gamma\!=\!\ri f_0 \mathcal{Z}/(\pi \mathcal{Z}_q)$. Here $f_0$ corresponds to the fundamental oscillation frequency and $\mathcal{Z}_q$ is the quartz shear-wave impedance, a known material constant \cite{review21}. In liquids, in contrast to vacuum where the adsorbed material only alters the inertial mass of the resonator contributing to the frequency shift, according to the Sauerbrey equation, the discrete adsorbates modify the viscous shear stress applied onto the sensor, contributing to the shifts in both the resonant frequency and the bandwidth (the latter is unaltered in vacuum). 

In the absence of adsorbates, the horizontal small-amplitude harmonic oscillations of the sensor located at $z\!=\!0$ with velocity $v_0\hat{\bm x} \cos{\omega t}$ create unidirectional oscillatory flow of the viscous liquid of viscosity $\eta$ and density $\rho$ occupying the half-space $z\!>\!0$ with velocity given by the real part of $v_0 \hat{\bm x} \re^{-z/\delta} \re^{-\ri(\omega t-z/\delta)}$ (e.g., see \cite{LL}, also known as Stokes' second problem). The flow disturbance propagates upward as the transverse wave attenuated by the exponential factor scaled with the \emph{viscous penetration depth}, $\delta=(2\nu/\omega)^{1/2}$, where  $\nu\!=\!\eta/\rho$ is the kinematic viscosity of the solvent (see Fig.~\ref{fig1}). At the fundamental frequency of oscillations $f_0\!=\!\omega/2\pi\!\simeq\!5$~MHz in aqueous solutions the penetration depth $\delta\!\simeq\!250$~nm. Computing the shear stress at the resonator, $\sigma_{xz}=\eta\, (\partial u_x/\partial z)_{z=0}$, and dividing by the surface velocity, yields the impedance ${\mathcal Z}\!=\!(\ri-1)\, \eta/\delta$ \cite{qcm-liquid85}, corresponding to a negative frequency and positive bandwidth shifts (as compared to unloaded resonator oscillating in vacuum). Obviously, the particles in the vicinity of the resonator would perturb this flow and modify the shear stress applied onto the sensor. The contribution to impedance due to the flow disturbance is \emph{fluid-mediated}, i.e., it takes place for both adsorbed and freely suspended particles, as it does not require a physical contact of the particle with the resonator. For the adsorbed particle in physical contact with the sensor, however, there is another hydrodynamic contribution to the impedance due to the force exerted on the particle by the liquid and transmitted to the resonator via contact. 
 
The prior works applied a variety of numerical methods to account for the hydrodynamics and compute the perturbed viscous stress exerted on resonator due to an adsorbed particle. Various factors, such as particle shape and size, surface coverage, particle mobility and other factors were considered using Finite Element method (FEM) in the early works \cite{johan09,tellech09}, and later with Lattice Boltzmann method (LBM) \cite{lbm1,lbm2,lbm3,freq-lbm}, Immersed Boundary (IB) method \cite{ibm20,ibm24} and most recently the spectral methods \cite{spectr25a,spectr25b}. Although the modern numerical approaches are quite powerful, the complex interplay of various factors and uncertainty of physicochemical properties and/or parameters governing the contact dynamics, call for an analytical approach able to decipher the role of the hydrodynamic forces in QCM-D analysis of discrete adsorbates. The analytical approach is particularly relevant in the low surface coverage limit, where the frequency-dependent \emph{excess shear force} (in excess of that without a particle) due to a single adsorbate can be readily interpreted as impedance due to a sparse monolayer of adsorbates by superposition, provided that their mutual hydrodynamics interaction is negligible. We have previously studied the excess shear force due to distant freely suspended particle \cite{prf23}, while the frequency response due to adsorbed particle rigidly attached to the resonator was studied in Ref.~\cite{analchem24}. In the later case, the analytical progress was possible in the limit of a small adsorbate with radius $a\!\ll\!\delta$, for which we demonstrated that that the hydrodynamic force dominates over the inertial force, and the apparent mass derived from QCM independently of the overtone number is about order-of-magnitude larger than the Sauerbrey (inertial) mass. The theoretical results showed an excellent agreement with the results of experiments and frequency-domain LBM simulations (see \cite{freq-lbm}) for a wide range of particle sizes and oscillation frequencies. However, not in all cases the particles adsorb forming a stiff contact, there is an experimental evidence that in many settings the adsorbates maintain a lubricated contact with the resonator. In agreement with experimental findings, showing a deviation from theory for $a/\delta\!\gtrsim\!0.5$, some simple scaling arguments suggest that even when a strong adherence is expected, the probability of forming a rigid contact with the adsorption site diminishes with the oscillation frequency and particle size due to inertia and viscous forces \cite{analchem24}.           

While the distant-particle theory in Ref.~\cite{prf23} provides an accurate prediction for the excess force down to a minimal separation distance between surfaces $d$ of about one particle diameter, it cannot be applied in the limit of $d\!\ll\!a$. The numerical study of the particle dynamics and excess force in close proximity to the resonator is challenging due to large (logarithmically divergent) viscous forces and it requires accurate resolution of the flow and pressure in the thin liquid gap. In the present paper we study the limit of small separations and propose an approximate \emph{lubrication theory} which allows to determine the excess shear force due to a small particle suspended in a liquid in a close proximity to the resonator, i.e., $d\!\ll\!a\!\ll\!\delta$. We have previously demonstrated \cite{prf23} that in the limit of small separations $\epsilon\!=\!d/a\!\ll\!1$ and  $\epsilon\! \ll\! (\delta/a)^2$, the lubrication equations of the \emph{steady} Stokes flow apply in the leading and sub-leading orders of the asymptotic expansion of the solution in small $\epsilon$. However, not much progress was made in \cite{prf23} for this limit, beyond showing that at contact, $d\!\to\!0$, the ``viscous adhesion" takes place, whereas the particle rotation velocity vanishes and it tends to oscillate in-sync (i.e., with the same amplitude and phase) with the resonator. Many experimentally relevant settings involving QCM measurement of adsorption of nanoscale discrete matter correspond to the limit $a\!\ll\!\delta$, and as we demonstrate below, the adequate lubrication-type approximation can be derived with the help of classical theories of steady Stokes flows.

\section{Problem formulation}

\begin{figure}[t]
\includegraphics[width=1.\textwidth]{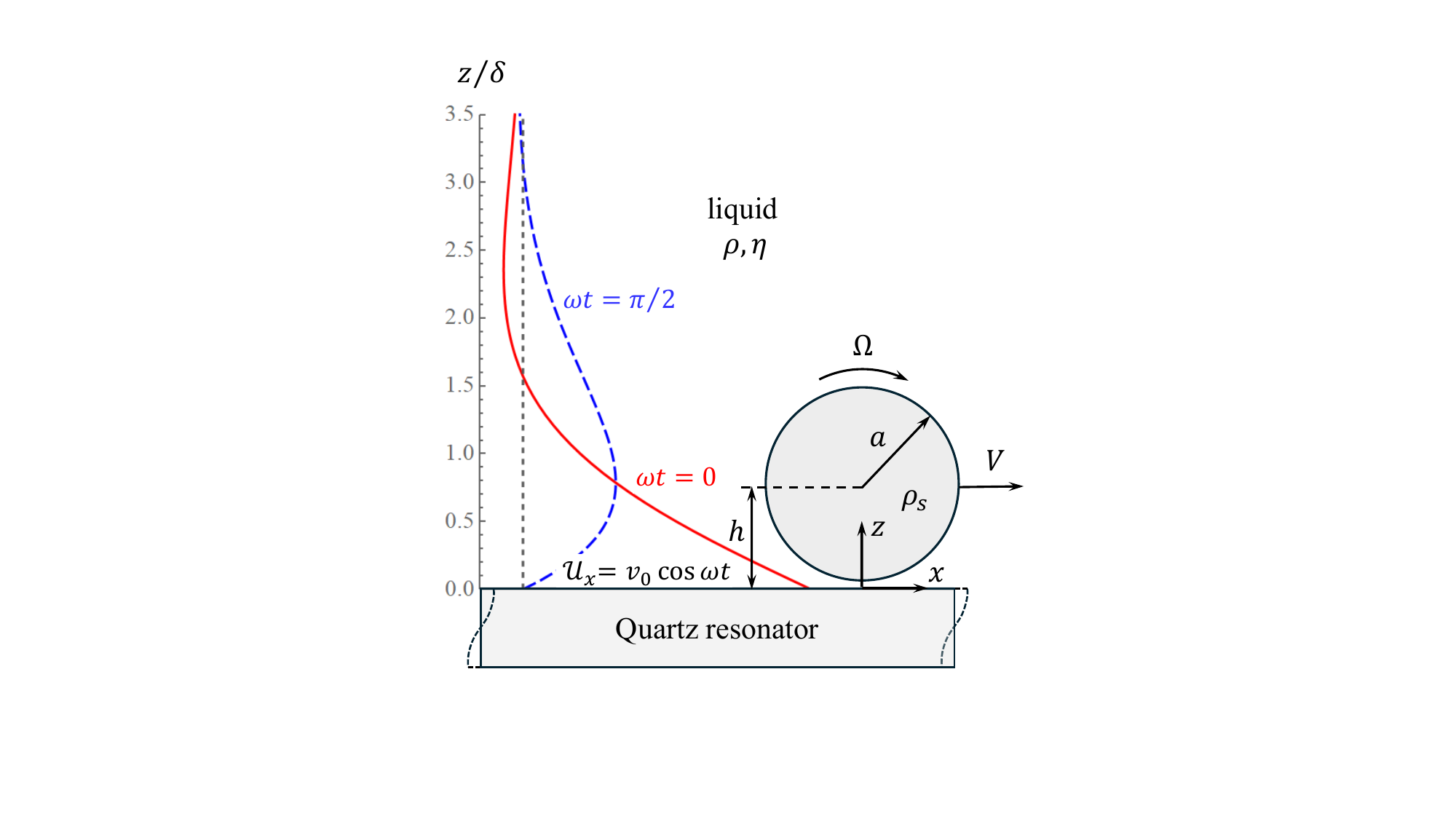}
\caption{Schematic illustration of the problem setting. A spherical particle of radius $a$ is positioned in the viscous liquid at distance $h$ above the horizontal plane at $z\!=\!0$, such that the lubrication gap thickness $d\!=\!h\!-\!a\!\ll\!a$; the plane is oscillating at MHz frequency with velocity $\mathcal U_x\!=\!v_0 \cos{\omega t}$. The undisturbed (i.e., in the absence of the particle) velocity profiles, ${\mathcal U}_x=v_0 \rRe[ \re^{-z/\delta} \re^{-\ri(\omega t-z/\delta)}]$, are shown at $\omega t\!=\!0$ (solid, red) and $\omega t\!=\!\pi/2$ (dashed, blue) vs. the scaled vertical distance $z/\delta$. The short-dashed vertical line stands for the zero value of the velocity. %\blue{[** add the arrows and the labels for the particle translation $V$ and rotation $\Omega$ velocities **]}
\label{fig1}}
\end{figure}

The spherical particle of radius $a$ and mass $m$ is assumed to be freely suspended in the incompressible viscous liquid occupying half-space $z\!>\!0$ with its center located at vertical elevation $z\!=\!h$ above a horizontally oscillating plate, such that $\epsilon\!=\!d/a \!\ll\! 1$, where $d\!=\!h-a$ is the minimal separation between the particle surface and the resonator (see the schematic drawing in Fig.~\ref{fig1}). The related problem was studied theoretically for a distant particle using the method of reflections and numerically for moderate separations in Ref.~\cite{prf23}, while the frequency response due to adsorbed particle (with stiff contact) was studied in Ref.~\cite{analchem24}. 

Below we consider the analytically trackable limit of a \emph{small} spherical particle suspended in a viscous solvent in \emph{close} proximity to the resonator, (i.e., $d\! \ll\! a\! \ll\! \delta$), while the minimal separation distance $d$ is held constant (by non-hydrodynamic, i.e., van der Waals and electrostatic, forces); the particle is free to translate parallel to the resonator and rotate about its center. 
We consider the setting where the fluid is set into motion by the periodic horizontal oscillations of the resonator located at $z\!=\!0$ along the $x$-axis with frequency $\omega$ and amplitude $v_0$ (see Fig.~\ref{fig1}). The shear stress that the particle above the oscillating plane exerts on it, depends on particle's motion. The time-periodic motion (i.e., translation along $x$-axis and rotation about $y$-axis) is not known \emph{a priori} and has to be determined self-consistently as part of the solution (see Ref.~\cite{prf23}). One can try to substantiate the conditions for lubricated contact by, e.g., examining the stiffness of the possible adhesive contact between the adsorbates and the sensor, by using the contact mechanics (see Ref.~\cite{analchem24} for details). However such analysis only provides a qualitative estimate, due to uncertainty in various physical parameters; it also does not take into account the sensor surface topography and charge distribution heterogeneity.

We also entirely neglect the Brownian transport of adsorbates, as it occurs on a much longer time scale compared to the characteristic time of the oscillations, $\tau_o\!=\!1/f$. The corresponding diffusion time $\tau_D\!=\!a^2/D_0$, where $D_0\!=\!k_B T/(6\pi\eta a)$ is the Stokes-Einstein self-diffusivity of the particle in the bulk. For $50$~nm particle in aqueous solution at $T\!=\!300$~K at the 3rd overtone, $f\!\approx\!15$~MHz, we estimate that $\tau_D/\tau_o\!\simeq\!1070$, indicating that the two time scales are well separated. Using the hindered diffusivity near the no-slip wall instead of $D_0$, would result in even higher value of $\tau_D$, further increasing the time scale separation. Moreover, the particles are typically bound to the surface by the electrostatic (or/and van der Waals) forces and its vertical transport is hindered, while lateral transport should not affect the measurements of the average stress.

Assuming small amplitude of the oscillations, $v_0/\omega \ll a$, to the leading approximation the flow $\bm {\mathcal U}$ and the pressure $P$ are governed by the unsteady Stokes equations
\begin{eqnarray}&&\!\!\!\!\!\!\!\!\!\!\!\!\!
\partial_t\, \bm {\mathcal U}\!=\!-\rho^{-1}\nabla P\!+\!\nu \nabla^2 \bm {\mathcal U},\ \ \nabla\cdot\bm {\mathcal U}\!=\!0,
\nonumber\\&&\!\!\!\!\!\!\!\!\!\!\!\!\! \bm {\mathcal U}(z\!=\!0) = v_0 \hat{\bm x} \cos{\omega t} ,\ \ \bm{\mathcal U}(r\!=\!a)=V \hat{\bm x}+ \Omega\, \hat{\bm y}\times \bm r\,, %\ \  \bm{\mathcal U}(z\!\to\!\infty) = 0\,, 
\label{fud}
\end{eqnarray}
where $V$ and $\Omega$ stand for the particle's linear and angular velocities, respectively, $\rho$ and $\nu\!=\!\eta/\rho$ are the density and the kinematic viscosity of the fluid and the radial distance $r=|\bm x-\bm x_c|$ is measured from the particle center located at $\bm x_c=(0, 0, h)$. We introduce dimensionless variables by normalizing the velocity with $v_0$, pressure with $\eta v_0/a$, time with $\omega^{-1}$ and distance with $a$. Notice that due to linearity of the unsteady Stokes equations, the plate oscillations result in particle displacement parallel to the plate (i.e., along the $x$-axis), and its rotation about the $y$-axis. The dimensionless (complex) flow field $\bm u$ and pressure $p$ defined via $\bm {\mathcal U}=v_0\rRe[\re^{-\ri \omega t} \bm u]$ and $P=\eta v_0\rRe[\re^{-\ri \omega t} p]/a$, where $\rRe$ stands for the real part, satisfy
\begin{eqnarray}&&\!\!\!\!\!\!\!\!\!\!\!\!\!
\lambda^2 \bm u\!=\!-\nabla p\!+\!\nabla^2\bm u,\ \ \nabla\cdot\bm u\!=\!0,
\nonumber\\&&\!\!\!\!\!\!\!\!\!\!\!\!\! \bm u(z=0)= \hat{\bm x},\ \ \bm u(r=1)=V \hat{\bm x}+\Omega\, \hat{\bm y}\times \hat{\bm r}\,, \label{fdon}
\end{eqnarray}
where $\lambda^2\!=\!-\ri a^2\omega/\nu=-2\ri(a/\delta)^2$, the dimensionless (complex) particle linear and angular velocity amplitudes, $V$ and $\Omega$, scaled with $v_0$ and $v_0/a$, respectively, which are governed by the Newton's force and torque balance:
\begin{eqnarray}&&\!\!\!\!\!\!\!\!\!\!\!\!\!
\lambda^2 \xi V\!=\! \oint_{r=1}\!\! \sigma_{xk} n_k dS,\label{force}\\
&&\!\!\!\!\!\!\!\!\!\!\!\!\! \frac{2}{5} \lambda^2   \xi \Omega \!= \!\oint_{r=1}\!\! \left[(z-h)\sigma_{xk}\!-\!x\sigma_{zk}\right] n_k dS,\label{torque}
\end{eqnarray}
where the parameter $\xi\!=\!m/(\rho a^3)$ and the repeated index $k$ implies summation over $k=\{x,y,z\}$. The real-valued (dimensionless) particle velocities are retrieved as $\rRe[V \re^{-\ri t}]$ and $\rRe[\Omega \re^{-\ri t}]$. From now on, unless explicitly stated otherwise, we treat all  variable as dimensionless.

\section{Excess shear force due to a freely suspended particle}

In the absence of a particle, the solution of Eqs.~(\ref{fdon}) is given by $\bm u_0=\re^{-\lambda z}\hat{\bm x}$, where $\lambda\!=\!(1-\ri)\,(a/\delta)$, and $p_0\!=\!0$ (see, e.g., \cite{LL}). When the particle is present, no analytical solution of Eqs.~(\ref{fdon}) is readily available. The stress tensor components corresponding to $\{\bm u, p\}$ in Eqs.~(\ref{fdon}) are given by $\sigma_{ik}\!\equiv\! - p\delta_{ik}\!+\!\partial_k u_i\!+\!\partial_i u_k$. In absence of the particle this tensor has only $xz$ component, which at the plate is equal to $-\lambda$. Our aim is determining the $x$-component of the dimensionless complex \emph{excess} shear force, $F$ (in excess of $-\lambda$) scaled with $\eta a v_0$ exerted on the oscillating plate due to presence of a freely suspended in a liquid particle located above the resonator, defined by
\begin{eqnarray}&&\!\!\!\!\!\!\!\!\!\!\!\!\!
F\! \equiv\!\int_{z=0}\! \left(\sigma_{xz}\!+\!\lambda\right) dx dy. \label{ff0}
\end{eqnarray}
Notice that for low particle surface coverage $\theta\!=\!{\tilde n}\pi a^2\!\ll\!1$, where $\tilde{n}$ is the particle areal number density, the mutual hydrodynamic interactions between particles can be neglected, and scaled excess shear force $F$ in Eq.~(\ref{ff0}) due to a \emph{single} particle multiplied by the total number of particles $N$ and divided by the surface area of the resonator $A$ becomes the dimensionless impedance, ${\!\mathcal Z}/(\eta a \tilde n)$ due to a sparse layer of particles probed by the QCM-D. 
The excess shear force due to a freely suspended \emph{distant} particle was studied theoretically in \cite{prf23} using the method of reflections and a close agreement with the numerical results was found for a wide range of particle sizes and separation distances down to a few particle radii. However, in the limit of vanishing proximity to the resonator, $\epsilon \!=\! d/a\! \to 0$, the theoretical approach becomes inaccurate, while the numerical scheme experiences difficulties due to confinement and high viscous forces. It is, however, possible to compute $F$ in the asymptotic limit of small particles for $|\lambda| \!\ll\! 1$. For that we shall use the auxiliary problem for flow perturbation, $\bm u'=\bm u-\re^{-\lambda z}\hat{\bm x}$, which satisfies:
\begin{eqnarray}&&\!\!\!\!\!\!\!\!\!\!\!\!\!
\lambda^2 \bm u'\!=\!-\nabla p\!+\!\nabla^2\bm u',\ \ \nabla\cdot\bm u'\!=\!0,
\nonumber\\&&\!\!\!\!\!\!\!\!\!\!\!\!\! \bm u'(z=0)= 0,\ \ \bm u'(r=1)=(V-\re^{-\lambda z})\hat{\bm x}+\Omega\, \hat{\bm y}\times \hat{\bm r}. \label{perturb}
\end{eqnarray}
The perturbed stress tensor $\sigma_{ik}'$ corresponding to $\{\bm u', p\}$ in Eq.~\ref{perturb} obeys $\lambda^2 u_i'\!=\!\partial_k \sigma'_{ik}$ and can be written as
\begin{eqnarray}\!\!\!\!\!\!\!\!\!\!\!\!\!
\sigma_{ik}'\!=\!\sigma_{ik}+\left(\delta_{ix}\delta_{kz}+\delta_{iz}\delta_{kx}\right)\lambda \re^{-\lambda z}\,.\label{stress}
\end{eqnarray}
Thus the excess shear force $F$ in Eq.~\ref{ff0} can then be written as
\begin{eqnarray}&&\!\!\!\!\!\!\!\!\!\!\!\!\!
F\!=\!\int_{z=0} \sigma_{xz}' dxdy=\!\int_{z=0} \partial_z u'_x dx dy. \label{fofc}
\end{eqnarray}

We shall also use the auxiliary problem of the flow generated in the vicinity of rigidly attached particle \cite{analchem24}, oscillating with the resonator without rotation:
\begin{eqnarray}&&\!\!\!\!\!\!\!\!\!\!\!\!\!
\lambda^2 \widehat{\bm v}\!=\!-\nabla \widehat p\!+\!\nabla^2 \widehat{\bm v},\ \ \nabla\cdot \widehat{\bm v}\!=\!0, \nonumber\\
&&\!\!\!\!\!\!\!\!\!\!\!\!\! \widehat{\bm v}(z\!=\!0)= \hat{\bm x},\ \ \widehat{\bm v}(r\!=\!1)=\hat{\bm x}. \label{adsorbed}
\end{eqnarray}

Applying the Lorentz reciprocity to the flow perturbation in Eqs.~\ref{perturb} and the \emph{dual} flow (hatted variabes) in Eqs.~(\ref{adsorbed}), satisfying $\lambda^2{\widehat v}_i=\partial_k{\widehat \sigma}_{ik}$, we have:
\begin{eqnarray}\!\!\!\!\!\!\!\!\!\!\!\!\!
\frac{\partial ({\widehat v}_i\sigma_{ik}')}{\partial x_k}= \frac{\partial (u_i'{\widehat \sigma}_{ik})}{\partial x_k}. \label{lorentz}
\end{eqnarray}
We integrate the Eq.~(\ref{lorentz}) over the volume of the semi-infinite domain above the resonator and apply the divergence theorem. The LHS of Eq.~\ref{lorentz} yields:
\bea
&& -\int_{z=0} {\widehat v}_i\sigma_{ik}' n_k dS - \oint_{r=1} {\widehat v}_i \sigma_{ik}' n_k dS= \nonumber \\
&& -F-\oint_{r=1} \sigma_{ik}' n_k dS= -F-\lambda^2\xi V-{\mathcal F}(\lambda)\,
\label{lhs}
\eea
where $ {\mathcal F}(\lambda)=4\pi \re^{-\lambda h} (\sinh{\lambda}\!-\!\lambda \cosh{\lambda})/\lambda$ and where we disregarded the integral at infinity due to fast decay of the flow disturbance and the corresponding stress (over the surface of half-sphere of infinitely large radius) and used the boundary condition 
$\widehat{\bm v}\!=\! \hat{\bm x}$ at $z\!=\!0$ and $r\!=\!1$; in the second equality in (\ref{lhs}) we substituted the perturbed stress from (\ref{stress}) and used the force balance (\ref{force}).    

The RHS of Eq.~(\ref{lorentz}) gives:
\bea
&& -\int_{z=0} u_i' \widehat{\sigma}_{ik} n_k dS - \oint_{r=1} u_i' \widehat{\sigma}_{ik} n_k dS= -V\! \underbrace{\oint_{r=1} \widehat{\sigma}_{xk} n_k dS}_{I_1} \nonumber \\
&&  - \Omega \underbrace{\epsilon_{iyk} \oint_{r=1} x_k  \widehat{\sigma}_{ij} n_j dS}_{I_2}+ \underbrace{\oint_{r=1} \re^{-\lambda z}  \widehat{\sigma}_{xk} n_k dS}_{I_3}\,, \label{rhs}
\eea
where we used the boundary condition $\bm u'\!=\!0$ at $z\!=\!0$ and disregarded over the half-sphere at infinity. Evaluating each of the three integral terms in (\ref{rhs}) we find that 
\be
I_1\equiv F_c'=\oint_{r=1} \widehat{\sigma}_{xk}' n_k dS-{\mathcal F}(\lambda) \,, \label{i1}
\ee
is (the hydrodynamic part of) the \emph{contact force} $F_c'$ in case of an adsorbed particle \cite{analchem24} and where $\widehat{\sigma}_{xk}'$ is the corresponding stress disturbance, $\widehat{\sigma}_{ik}'\!=\!\widehat{\sigma}_{ik}+\left(\delta_{ix}\delta_{kz}+\delta_{iz}\delta_{kx}\right)\lambda \re^{-\lambda z}$. Analogously, we find that
\bea
I_2 &=& \epsilon_{yki} \oint_{r=1} x_k  \widehat{\sigma}_{ij} n_j dS \equiv L_c \nonumber \\
&& %\epsilon_{yki} \oint_{r=1} x_k  \widehat{f}_{i}' dS 
=\oint_{r=1} \left[(z-h)\widehat{\sigma}_{xk}'\!-\!x\widehat{\sigma}_{zk}'\right] n_k dS - \mathcal{G}(\lambda) \,. \label{i2}
\eea
is the ($y$-component of the) \emph{contact torque} $L_c$, which could also be computed integrating the perturbed traction $\widehat{\sigma}_{ij}'\!n_j$ at the surface of the adsorbed particle \cite{analchem24}, where ${\mathcal G}(\lambda)\!=\! 4\pi \re^{-\lambda h}[\sinh{\lambda} +
3 \lambda^{-2}(\sinh{\lambda}-\lambda\cosh{\lambda})]$.
The last integral yields:
\be
I_3\!=\!\oint_{r=1} \re^{-\lambda z}  \widehat{\sigma}_{xk}' n_k dS - \mathcal{H}(\lambda)\,, \label{i3}
\ee
where ${\mathcal H}=\pi \lambda^{-1} \re^{-2\lambda h} (\sinh{2\lambda}-2\lambda \cosh{2\lambda})$. \\

Collecting all terms and equating the RHS to the LHS, we readily obtain the expression for the excess force $F$ due to a freely suspended particle as:
\be
F=V F_c'+\Omega L_c-\oint_{r=1} \re^{-\lambda z}  \widehat{\sigma}_{xk}' n_k dS+{\mathcal H}(\lambda)-{\mathcal F}(\lambda)-\lambda^2 \xi V. \label{Ffull}
\ee
From Eq.~(8) in Ref.~\cite{analchem24} it follows that that the three terms in (\ref{Ffull}) are exactly equal to the fluid-mediated \emph{excess} shear force exerted on the oscillating sensor due to presence of an \emph{adsorbed} particle,  $F_a\! \equiv\!\int_{z=0} \widehat{\sigma}'_{xz} dx dy$: 
\be
F_a\!=\! %\int_{z=0}\! \widehat{\sigma}_{xz}'\! dx dy\!=\!
-\oint_{r=1} \re^{-\lambda z}  \widehat{\sigma}_{xk}' n_k dS+{\mathcal H}(\lambda)-{\mathcal F}(\lambda)\,. \label{Fa}
\ee
Therefore, using (\ref{Fa}) in (\ref{Ffull}), the excess shear force due to a due to a freely suspended particle in Eq.~(\ref{Ffull}) reduces to a particularly compact representation: 
\be
F\!=\!V F_c+\Omega L_c+F_a\,\label{Fcompact}
\ee
where $F_c=F_c'-\lambda^2 \xi$ is the net contact force (including the particle inertia, see Ref.~\cite{analchem24}). Recall that we have previously used an alternative representation (see Eq.~\ref{Faltrn} below) for $F$, derived via the use of complementary problems of an oscillating plane in presence of a \emph{stationary} spherical particle, and a spherical particle undergoing prescribed oscillatory translation and rotation with unit velocity, above a stationary plane \cite{prf23}. For the purpose of the asymptotic small-$\lambda$ analysis below, we shall use the representation (\ref{Fcompact}), as $F_c$, $L_c$ and $F_a$ have been previously determined for particle oscillating \emph{in contact} (i.e., for $h\!=\!1$) with the plane, in Ref.~\cite{analchem24}. Therefore, computation of the excess shear force $F$ in Eq.~(\ref{Fcompact}) reduces to determining the particle velocities, $V$ and $\Omega$, which depend on $\epsilon$, $\xi$ and $\lambda$. In the limit of contact, $h \!\to\! 1$, we have $V\! \to\! 1$ and $\Omega \! \to \! 0$, such that the excess shear force in Eq.~(\ref{Fcompact}) goes to that due to an adsorbed particle with a rigid contact, $F\!\to\!F_a+F_c'-\lambda^2 \xi$. While $F_c'$, $F_a$ and $L_c$ in (\ref{Fcompact}) at contact were computed numerically {and tabulated vs. $a/\delta$ \cite{analchem24}, numerical computation of the velocities $V$ and $\Omega$ of a freely suspended particle is, however, challenging due to their logarithmic divergence at $h \!\to\! 1$. 

Notice that upon entirely neglecting the hydrodynamics in Eq.~(\ref{Fcompact}), i.e., for $F_a\!=\!F_c'\!=\!L_c\!=\!0$, for  the net shear excess force due to inertial mass of the rigidly adsorbed particle (i.e., for $V\!=\!1$, $\Omega\!=0\!$) reads $F\!=\!-\lambda^2\xi\!=\!-(4\pi/3)(\rho_s/\rho)\lambda^2\!=\!\ri m\omega/\eta a$, which is equivalent, upon summing up contributions of individual adsorbates, to the (dimensional) impedance ${\mathcal Z}=\ri\omega {\tilde n} m$. Substituting $\omega=2\pi n f_0$ and using the small-load approximation for the impedance, $\Delta f-\ri \Delta\Gamma\!=\!\ri f_0 \mathcal{Z}/(\pi \mathcal{Z}_q)$ \cite{review21}, we readily arrive at the seminal Sauerbrey equation \cite{sauerbrey59} relating the adsorbed mass to the shift of resonant frequency:
\begin{equation}
\frac{\Delta f_S}{f_0} \simeq \ri \frac{\mathcal Z}{\pi {\mathcal Z}_q}=-n\, \frac{{\tilde n} m}{m_q}\;,\label{sauerbrey}
\end{equation}
where $m_q\!=\!\mathcal{Z}_q/(2 f_0)$ and $m {\tilde n}$ is the areal mass density (both have units of mass per unit area).

\subsection{Small weakly inertial particles}
The analytical progress can be made in the limit of small particles, $|\lambda|\!\ll\!1$. Expanding the exponent under the integral sign for small $\lambda$ in Eq.~(\ref{Fa}) and using that $F_c'\!=\!\oint_{r=1} \widehat{\sigma}_{xk}' n_k dS-{\mathcal F}(\lambda)$ we readily find from Eq.~(\ref{Fcompact}) that
\be
F\approx (V-1) F_c'+\Omega L_c+\lambda \oint_{r=1} z \widehat{\sigma}_{xk}' n_k dS -\lambda^2 \xi V+\ldots\,, \label{F}
\ee
where the dots stand for the neglected terms estimated as ${\mathcal O}(|\lambda|^3)$ due to the fact that $\widehat{\sigma}_{xk}' ={\mathcal O}(|\lambda|)$ (see \cite{analchem24}) and ${\mathcal H}-2{\mathcal F}={\mathcal O}(|\lambda|^3)$. Notice that for an adsorbed (i.e., rigidly attached) particle ($V\!=\!1$, $\Omega\!=\!0$) the result in (\ref{F}) reduces to Eq.~(19) in Ref.~\cite{analchem24}. 

In the small-particle limit, $|\lambda|\ll 1$, the steady Stokes equations hold to the first approximation, as the unsteady term $\lambda^2\bm{u}'$ in Eqs.~(\ref{perturb}) produces $o(|\lambda^2|)$ corrections in the solution. We next expand the perturbed flow and pressure (or stress) in Eqs.~(\ref{perturb}) in the power series in $\lambda$ as $\bm u'=\lambda  \bm u_1'+\lambda^2 \bm u_2' + \ldots$, $p=\lambda p_1'+\lambda^2 p_2'+ \ldots$, such that at the leading order we have:
\begin{eqnarray}&&\!\!\!\!\!\!\!\!\!\!\!\!\!
-\nabla p_1'\!+\!\nabla^2\bm u_1'\!=\!0,\ \ \nabla\cdot\bm u_1'\!=\!0\,,
\nonumber\\&&\!\!\!\!\!\!\!\!\!\!\!\!\! \bm u_1'(z\!=\!0)=0,\ \ \bm u_1'(r\!=\!1)=U_1 \hat{\bm x}\!+\!\Omega_1\, \hat{\bm y}\!\times\! \hat{\bm r}\!+\!z\hat{\bm x}\,, \label{brenner}
\end{eqnarray}
where we assumed $V-1=\lambda U_1+\ldots$ and $\Omega=\lambda \Omega_1+\ldots$ to be the leading-order [i.e., ${\mathcal O}(|\lambda|)$] linear and angular velocities of the particle. It also follows from Eqs.~(\ref{force})--(\ref{torque}) that provided that particle is not too heavy (i.e., $\lambda^2\xi \ll \lambda$, or $\xi\!\ll |\lambda|^{-1}$, which for neutrally buoyant particles gives $a/\delta \ll 0.17$) to the leading approximation the particle can be considered force- and torque-free,
\[
\oint_{r=1}\!\! \sigma_{xk}^{\prime\,(1)} n_k dS\!=0,\qquad 
\oint_{r=1} \! \left[(z-h)\sigma_{xk}^{\prime\,(1)}\!-\!x \sigma_{zk}^{\prime\,(1)}\right] n_k dS\!=\!0 \,.
\]
The Stokes problem in Eqs.~(\ref{brenner}) for a motion of force- and torque-free particle in a simple-shear flow near a plain rigid wall was solved in Ref.~\cite{brenner67}. The problem formulation in Ref.~\cite{brenner67} differs from that in Eqs.~(\ref{brenner}) by the last term, $-Sz\hat{\bm x}$. Using the resulting expressions for the particle velocities $U_1$ and $\Omega_1$ in Eqs.~(4.11) and (4.12) in \cite{brenner67} derived using lubrication theory for the particle in close proximity to the wall, $\epsilon\! \ll \! 1$, and substituting $S=-1$ and $h=1+\epsilon$ the resulting expressions for the particle linear and angular velocities read:
\be
V \simeq 1-\frac{0.7431 (1+\epsilon) \lambda}{0.6376+0.200 \ln{\epsilon^{-1}}}\,, \qquad \Omega \simeq -\frac{0.4218\lambda}{0.6376+0.200 \ln{\epsilon^{-1}}}\,. \label{U1Omega1}
\ee
As expected, $V-1,\,\Omega\!\to\! 0$ at $\epsilon\!\to\!0$ logarithmically slow.    

The asymptotic (small-$\lambda$) expressions for $F_c'$ and $L_c$ were obtained in \cite{analchem24} (see also \cite{arxiv24}), employing the revisited analysis of O'Neill \cite{oneill68}, and they read
\be
F_c'= -6\pi\lambda f\,,\quad L_c= -4\pi \lambda g\, \label{FcLc}
\ee  
where $f\! \simeq\! 1.7005$ and $g\!\simeq \!0.944$. \\

The third (integral) term in Eq.~(\ref{F}) was previously computed in \cite{analchem24} using the revisited O'Neill's analysis in Ref.~\cite{oneill68}, 
\be
\oint_{r=1} z \widehat{\sigma}_{xk}' n_k dS = -40.159\,\lambda \,. \label{z-moment}
\ee
Substituting the results in Eqs.~(\ref{U1Omega1}), (\ref{FcLc}) and (\ref{z-moment}) into Eq.~(\ref{F}) we obtain the asymptotic expression for the excess shear force (scaled with $\eta a v_0$) due to a freely suspended \emph{small weakly inertial} particle at close proximity to the resonator (i.e., in the limit $d\!\ll\! a \!\ll\!\delta$):
\be
F\simeq \left[\frac{7.582 (1+\epsilon)+ 1.593}{0.6376+0.200 \ln{\epsilon^{-1}}} \right]\pi \lambda^2-(40.159+\xi)\lambda^2\,. 
%F\simeq \left[\frac{9.175 \pi }{0.6376+0.200 \ln{\epsilon^{-1}}} - (40.159+\xi)\right]\lambda^2\,. 
\label{close}
\ee   
Notice that since $\lambda^2\!=\!-\ri a^2\omega /\nu$ is purely imaginary, the leading contribution to the real part of $F$ reduces to ${\mathcal O}(|\lambda|^3)$, implying that for small particles the dissipation shift is expected to be smaller in comparison to the frequency shift by a factor $\propto\!|\lambda|$, as for the adsorbed particle \cite{analchem24}. The real part of $F$ would require higher order analysis. Recall that the numerator in the square brackets in (\ref{close}) is a sum of two contributions -- one due to particle translation ($\simeq\!7.582\pi$) and the other one, about five times smaller ($\simeq\!1.593\pi$), being contributed by the particle rotation. It is worth pointing out, that the second term in the r.h.s. of Eq.~(\ref{close}) is equal to the expression for the excess shear force due to an adsorbate with a stiff contact (see Eq.~19 in Ref.~\cite{analchem24}), implying that in the limit of contact at $\epsilon\!\to\!0$, the first term in the r.h.s. of Eq.~(\ref{close}) vanishes (logarithmically slow) and the response tends to that due to an adsorbate with a stiff contact.

\subsection{Small particles of arbitrary density}
If particle are small, with $|\lambda|\ll 1$, but arbitrarily heavy, i.e., $|\lambda|\xi \geq {\mathcal O}(1)$ (e.g., Pt nanoparticles with $\rho_s/\rho\!=\!21.4$), then one cannot neglect the inertia terms in the force- and torque balances. For $|\lambda|\ll 1$ Eqs.~(\ref{perturb}) yet reduces to the steady Stokes equations to the leading approximation with an error ${\mathcal O}(|\lambda|^2)$:
\begin{eqnarray}
&&\!\!\!\!\!\! -\nabla p'\!+\!\nabla^2\bm u'\!=\!0,\ \ \nabla\cdot\bm u'\!=\!0, \nonumber\\
&& \bm u'(z=0)= 0,\ \nonumber \\
&& \bm u'(r=1)=(V-1+\lambda z)\hat{\bm x}+\Omega\, \hat{\bm y}\times \hat{\bm r}. \label{perturb1}
\end{eqnarray}
The force- and the torque balances for the perturbed stress $\sigma_{ik}^\prime$ corresponding to $\{u'_i, p'\}$ in Eq.~(\ref{perturb1}) take the form:
\bea
&& \oint_{r=1}\!\! \sigma_{xk}^{\prime} n_k dS\!=\lambda^2\xi V, \nonumber \\ 
&& \oint_{r=1} \! \left[(z-h)\sigma_{xk}^{\prime}\!-\!x \sigma_{zk}^{\prime}\right] n_k dS\!=\!\frac{2}{5} \lambda^2  \xi \Omega\,. \nonumber
\eea
The solution to the linear Eqs.~(\ref{perturb1}) can be constructed by a superposition of (i) a solution of a particle moving with translation velocity $(\widetilde{V}\!-\!1)$ and angular velocity $\widetilde{\Omega}$ when driven near a wall by a finite force $\lambda^2\xi \widetilde{V}$ and torque $\frac{2}{5}\lambda^2 \xi \widetilde{\Omega}$ and (ii) a solution for a force- and torque-free particle translating and rotating in a linear wall-bounded shear flow $-\lambda z\hat{\bm x}$ with velocities $\lambda U_1$ and $\lambda\Omega_1$ obtained in Eqs.~(\ref{U1Omega1}). The resulting velocities $V=\widetilde{V}+\lambda U_1$ and $\Omega=\widetilde{\Omega}+\lambda \Omega_1$ will be used to compute the excess force in Eq.~(\ref{F}).

The problem (i) has the following form:
\bea&&\!\!\!\!\!\!\!\!\!\!\!\!\!
\left(\begin{array}{cc}
\lambda^2\xi {\widetilde V}\\
\frac{2}{5}\lambda^2\xi {\widetilde \Omega}  \end{array}\right)=\bm{\mathcal R} \left(\begin{array}{cc}
{\widetilde V}-1 \\
{\widetilde \Omega}  \end{array}\right), \label{veliner}
\eea
where $\bm{\mathcal R}$ is a symmetric $2\!\times\!2$ \emph{resistance} matrix, given to the leading approximation for $\epsilon\!\ll\!1$ by ${\mathcal R}_{11}\!=\!{\mathcal R}_{22}\!=\!-\frac{16\pi}{5}\ln{\epsilon^{-1}}+{\mathcal O}(1)$ and
${\mathcal R}_{12}\!=\!{\mathcal R}_{12}\!=\!\frac{4\pi}{5}\ln{\epsilon^{-1}}+{\mathcal O}(1)$
\cite{kk91}.
The omitted ${\mathcal O}(1)$ terms in the individual components ${\mathcal R}_{ik}$ are found from matching the lubrication solution to the outer (e.g., numerical) solution; they may not be negligible, unless the dimensionless proximity $\epsilon$ is exponentially small, $\ln{\epsilon^{-1}}\!\gg \!1$. More accurate results read 
\be
\bm {\mathcal R}= \,
\left( \begin{array}{cc}
-\frac{16\pi}{5}\ln{\epsilon^{-1}}-6\pi c_1  & \frac{4\pi}{5}\ln{\epsilon^{-1}}-6\pi c_2 \\
\frac{4\pi}{5}\ln{\epsilon^{-1}}-6\pi c_2 & -\frac{16\pi}{5}\ln{\epsilon^{-1}}-8\pi c_3
\end{array} \right)+{\mathcal O}(\epsilon\ln{\epsilon^{-1}})\,, \label{R1}
\ee 
where $c_1\!\simeq \!0.9543$, $c_2\!\simeq \!0.2573$ and $c_3\!\simeq\!0.3709$ (see, e.g., \cite{cooley68}, \cite{stew67}). 

The solution of the linear system of Eqs.~\ref{veliner} reads:
\bea
{\widetilde V}&=&\frac{{\mathcal R}_{12}^2-{\mathcal R}_{11} ({\mathcal R}_{22}-\frac{2}{5} \lambda ^2 \xi)}{\left(\lambda ^2 \xi -{\mathcal R}_{11}\right) \left({\mathcal R}_{22}-\frac{2}{5} \lambda ^2 \xi \right)+{\mathcal R}_{12}^2}\, \label{Viner} \\
{\widetilde \Omega} &=& \frac{\lambda ^2 \xi  {\mathcal R}_{12}}{\left(\lambda ^2 \xi -{\mathcal R}_{11}\right) \left({\mathcal R}_{22}-\frac{2}{5} \lambda ^2 \xi \right)+{\mathcal R}_{12}^2}\,. \label{Ominer}
\eea
In the limit of vanishing proximity $\epsilon\!\rightarrow \!0$ (or vanishingly small particle, $|\lambda| \rightarrow 0$) we have ${\widetilde V}\rightarrow 1$ and ${\widetilde \Omega}\rightarrow 0$ for a finite $\xi$, while in the limit of high particle inertia, $\xi\! \rightarrow \!\infty$, the particle's motion vanishes as ${\widetilde V}\!\sim\! -\frac{{\mathcal R}_{11}}{\lambda^2\xi}$ and ${\widetilde \Omega} \!\sim\! -\frac{5}{2}\frac{{\mathcal R}_{12}}{\lambda^2\xi}$ in agreement with Eq.~25 in Ref.~\cite{prf23}. 

The resulting particle velocity then would read
\be
V \simeq  {\widetilde V}-\frac{0.7431 (1+\epsilon) \lambda}{0.6376+0.200 \ln{\epsilon^{-1}}}\,,\qquad 
\Omega \simeq {\widetilde \Omega} -\frac{0.4218\lambda}{0.6376+0.200 \ln{\epsilon^{-1}}}\,. \label{U2Omega2}
\ee  

Substituting the results in Eqs.~(\ref{U2Omega2}), (\ref{FcLc}) and (\ref{z-moment}) into Eq.~(\ref{F}) it can be shown that the expression for the dimensionless excess shear force $\widetilde{F}$ (scaled with $\eta a v_0$) due to a freely suspended small \emph{inertial} particle (i.e., $|\lambda|\! \ll \!1$ and arbitrary  $\xi$) at close proximity can be written as 
\be
\widetilde{F}=({\widetilde V}-1)F_c'+ {\widetilde \Omega}L_c+F \label{close1}
\ee 
where $F$ is given by Eq.~\ref{close}. 
Notice that for heavy (inertial) particles, we expect $\widetilde{F}$ in Eq.~(\ref{close1}) to be dominated by the first two terms contributing to both the real and the imaginary part of the expression. Since $({\widetilde V}\!-\!1)$ and  ${\widetilde \Omega}$ are both $\mathcal{O}\left(\frac{\lambda^2 \xi}{\ln{1/\epsilon}}\right)$ while $F_c'$ and $L_c$ are $\mathcal{O}\left(\frac{\lambda}{\ln{1/\epsilon}}\right)$ at $\epsilon\!\to\! 0$, then the condition for $\widetilde{F}$ to be dominated by the first two terms is $\xi\! \gg \! |\lambda|^{-1}$, which for, e.g., Pt nanoparticles ($\rho_s/\rho\!=\!21.45$) transforms into $a/\delta\!\gg\! 0.008$. In such case, Eq.~(\ref{close1}) can accurately predict not only the leading-order asymptotic behavior of the imaginary part of $\widetilde{F}$ (frequency shift) near contact, but also its real part (dissipation factor), which is not readily available for weakly inertial (e.g., neutrally buoyant) nanoparticles. %This is readily demonstrated in Fig.~\ref{fig3}(c), where the asymptotic prediction of both, the real and imaginary part of the impedance near contact in Eq.~(\ref{close1}) are in excellent agreement with the numerical results for Pt nanoparticles.

\section{Numerical solution}
We solve Eqs.(\ref{perturb}) numerically and compute the excess shear force in (\ref{fofc}) at arbitrary separation distance $h$, we apply the approach proposed in Ref.~\cite{prf23} which uses an alternative to Eq.(\ref{Fcompact}) representation of the excess shear force,  
\be
F=F_s+V{\mathcal A}+\Omega{\mathcal B}\,. \label{Faltrn}
\ee 
Here $F_s$ is the excess shear force acting on an oscillating plane in presence of a \emph{stationary} particle, while ${\mathcal A}$ and ${\mathcal B}$ are the resistance coefficients equal to the force and torque, respectively, exerted on the stationary plane at $z\!=\!0$ due to a prescribed oscillatory (translation and rotation) motion with unit velocity of a spherical probe. Therefore computation of $F_s$ requires the numerical solution of the auxiliary problem (i) for the complex flow velocity and pressure $\{\widehat{\bm v}, \widehat{p}\}$ satisfying the Stokes and continuity equations (\ref{adsorbed}) with the boundary conditions $\widehat{\bm v}\!=\!0$ at $z\!=\!0$ and $\widehat{\bm v}\!=\!-\re^{-\lambda z} \hat{\bm x}$ at $r\!=\!1$. Similarly, ${\mathcal A}$ and ${\mathcal B}$ are determined from solving the two complementary problems for the flow satisfying the same Eqs.~(\ref{adsorbed}) with the boundary conditions $\widehat{\bm v}\!=\!0$ for both at $z\!=\!0$ and (ii) $\widehat{\bm v}\!=\!\hat{\bm x}$ and (iii)  $\widehat{\bm v}\!=\!\hat{\bm y}\times \hat{\bm r}$ at the particle surface $r\!=\!1$.

The particle linear and angular velocities are readily obtained from the solution of the complementary problems (ii) and (iii) as \cite{prf23}:
\begin{eqnarray}&&\!\!\!\!\!\!\!\!\!\!\!\!\!
\left(\begin{array}{cc}
 V \\
\Omega  \end{array}\right)=-\frac{1}{\left({\mathcal R}_{11}-\lambda^2 \xi \right)\left({\mathcal R}_{22}-\frac{2}{5} \lambda ^2 \xi \right)-{\mathcal R}_{12}^2}
\left(\begin{array}{cc}
{\mathcal R}_{22}-\frac{2}{5} \lambda ^2 \xi   & -{\mathcal R}_{12} \\
-{\mathcal R}_{12} &  {\mathcal R}_{11}-\lambda^2 \xi \end{array}\right)\!\!\left(\begin{array}{cc}
{\mathcal A} \\
{\mathcal B} \end{array}\right)\,,\label{motion}
\end{eqnarray}
where ${\mathcal R}_{ij}$ are the components of the symmetric $2\!\times\!2$ complex resistance matrix, relating the $x$-component of the force and the $y$-component of the torque to the particle's oscillatory translation and angular velocities, 
\begin{eqnarray}&&\!\!\!\!\!\!\!\!\!\!\!\!\!
\left(\begin{array}{cc}
\widehat{F}\\
\widehat{L}  \end{array}\right)=\bm{\mathcal R} \left(\begin{array}{cc}
\widehat{V} \\
\widehat{\Omega}  \end{array}\right)\,. \label{resist}
\end{eqnarray}  
The components of $\bm{\mathcal R}$ are determined from the solution of the above complementary problems (ii) and (iii) along with $\mathcal A$ and $\mathcal B$. The auxiliary problems (i)--(iii) can be treated similarly using the cylindrical coordinates, $(\varrho, \phi, z)$, such that $x\!=\!\varrho \cos{\phi}$, $y\!=\!\varrho \sin{\phi}$, $z\!=\!z$, with its origin at $z\!=\!0$ with the $z$-axis passing through the center of the spherical particle (see Fig.~\ref{fig1}). In cylindrical coordinates the unsteady Stokes equation in (\ref{adsorbed}) for the pressure $\widehat{p}$ and the velocity $\widehat{\bm v}$ can be written as (e.g., \cite{prf23}):
\begin{eqnarray}&&\!\!\!\!\!\!\!\!\!\!\!\!\!
\lambda^2 {\widehat v}_{\varrho}\!=\!-\frac{\partial \widehat p}{\partial \varrho}\!+\!\nabla^2 {\widehat v}_{\varrho}-\frac{{\widehat v}_{\varrho}}{\varrho^2}-\frac{2}{\varrho^2} \frac{\partial {\widehat v}_\phi}{\partial \phi}\,, \nonumber \\
&& \!\!\!\!\!\!\!\!\!\!\!\!\!\lambda^2 {\widehat v}_{\phi}\!=\!-\frac{1}{\varrho}\frac{\partial \widehat p}{\partial \phi}\!+\!\nabla^2 {\widehat v}_{\phi}-\frac{{\widehat v}_{\phi}}{\varrho^2}+\frac{2}{\varrho^2}\frac{\partial {\widehat v}_{\varrho}}{\partial\phi}\,,\nonumber\\
&&\!\!\!\!\!\!\!\!\!\!\!\!\!  \lambda^2 {\widehat v}_z\!=\!-\frac{\partial \widehat p}{\partial z}\!+\!\nabla^2 {\widehat v}_z\,, \label{Stokes_cyl}
\end{eqnarray}
where
\begin{eqnarray}&&\!\!\!\!\!\!\!\!\!\!\!\!\!
\nabla^2=\frac{1}{\varrho}\frac{\partial}{\partial \varrho} \left(\varrho\frac{\partial}{\partial \varrho}\right)+\frac{1}{\varrho^2}\frac{\partial^2}{\partial \phi^2}+\frac{\partial^2}{\partial z^2}\,, \nonumber
\end{eqnarray} 
while the continuity equation takes the form
\be
\frac{1}{\varrho} \frac{\partial}{\partial\varrho} \left(\varrho {\widehat v}_\varrho \right)+
\frac{1}{\varrho} \frac{\partial {\widehat v}_\phi}{\partial\phi}+\frac{\partial {\widehat v}_z}{\partial z}=0\,. \label{cont}
\ee

The reduction to two spatial dimensions can be made by noting the dependence of the solution on the azimuthal angle $\phi$ via the ansatz (the same as for steady Stokes flow, see \cite{kk91}):
\begin{eqnarray}
&& {\widehat v}_{\varrho}={\mathcal U}(\varrho, z)\cos\phi, \ \ {\widehat v}_{\phi}={\mathcal V}(\varrho, z)\sin\phi,\nonumber\\
&& {\widehat v}_z= {\mathcal W}(\varrho, z)\cos\phi,\ \ {\widehat p}={\mathcal P}(\varrho, z)\cos\phi\,, \label{ansatz}
\end{eqnarray}
where ${\mathcal U}$, ${\mathcal V}$, ${\mathcal W}$ and ${\mathcal P}$ are functions of the radial, $\varrho$, and axial, $z$, cylindrical coordinates. Substituting the ansatz (\ref{ansatz}) into the Eqs.~(\ref{Stokes_cyl}) yields:
\begin{eqnarray}
&& \lambda^2 {\mathcal U}=-\partial_{\varrho} {\mathcal P}\!+\!L_0^2\, {\mathcal U}-\frac{2({\mathcal U}+{\mathcal V})}{\varrho^2},\nonumber\\
&&  \lambda^2  {\mathcal V}=\frac{{\mathcal P}}{\varrho}\!+\!L_0^2 {\mathcal V}-\frac{2({\mathcal U}+{\mathcal V})}{\varrho^2},\nonumber\\
&& \lambda^2  {\mathcal W}=-\partial_z {\mathcal P}\!+\!L_1^2 \mathcal W, \label{Stokes_cyl1}
\end{eqnarray}
where the operator $L_m^2$ is defined as
\[
L_m^2=\frac{\partial^2}{\partial \varrho^2}+\frac{1}{\varrho}\frac{\partial}{\partial \varrho}-\frac{m^2}{\varrho^2}+\frac{\partial^2}{\partial z^2}\,,
\]
while the continuity equation (\ref{cont}) reduces to
\be
\frac{\partial \mathcal U}{\partial \varrho}+\frac{{\mathcal V}+{\mathcal U}}{\varrho}+\frac{\partial \mathcal W}{\partial z}=0. \label{cont1}
\ee

The numerical solution for $\mathcal U$, $\mathcal V$, $\mathcal W$ and $\mathcal P$ in Eqs.~(\ref{Stokes_cyl1})--(\ref{cont1}) is performed in 
in the rectangular domain $0\le\varrho\!\le\!\varrho_{m},\; 0\!\le\! z\! \le\! z_\mathrm{m}$ with an exclusion of the half unit disk $\varrho^2+(z-h)^2\! \le\!1$ centered at $(0, h)$ representing the particle (see also \cite{prf23}).  The pressure $\mathcal P$ is set to a fixed (zero) value far from the particle at $z\!=\!z_\mathrm{m}$, $\varrho\!=\!\varrho_\mathrm{max}$. The boundary condition $\mathcal U \!=\!\mathcal V\!=\!\mathcal W\!=\!0$ is used at $\varrho\!=\!\varrho_\mathrm{m}$,  $z\!=\!0$ and $z\!=\!z_\mathrm{m}$. We set no-flux boundary condition at $\varrho\! =\!0$, while at the half-circle representing the particle surface the boundary condition depends on type of the problem (i.e., stationary, translating or rotating). In particular, for the problem (i) for determining $F_s$ we specify $\mathcal U\! =\! - \mathcal V\!=\! -\re^{-\lambda z}$ and $\mathcal W \!=\! 0$.
For the computation of $\mathcal A$ and ${\mathcal R}_{11}$, ${\mathcal R}_{12}$ in (ii) we set $\mathcal U \!=\! - \mathcal V\!=\!1$, $\mathcal W\! =\! 0$, while finding $\mathcal B$ and ${\mathcal R}_{22}$ in (iii) requires $\mathcal U\! =\!-\mathcal V\!=\!z-h$ and $\mathcal W\! =\! -\varrho$. We used the Finite Element method (FEM) implemented in {\it Mathematica} 12.0. The computation domain was defined by $\varrho_m\! =\! 15$, $z_m \!=\! 10+h$. The maximal mesh cell area was selected to be $0.025$ within the domain and the maximal side length of $0.008$ along the boundaries, resulting in $\sim\!50$k mesh cells overall. For particle at close proximity to the plane (i.e., for $1.002\!\leq\!h\!\leq\!2$) more accurate computations required the mesh with the maximal cell area of $0.01$ in the domain and the maximal side length of $0.003$ along the boundaries, resulting in overall $\sim\!90$k mesh cells in a smaller domain, $\varrho_m\! =\! 10$, $z_m \!=\! 10+h$. 

\section{Materials and Experimental Methods}

Positively charged polystyrene (PS) particles supplied by Invitrogen (Life Technologies Polska Sp.z.o.o., Warsaw, Poland) were used in the deposition kinetics measurements carried out by QCM. Particle density $\rho_s$ determined by the densitometry/dilution method was equal to $1.05$~g~cm$^{-3}$. Their diffusion coefficients, hydrodynamic diameters and zeta potentials at $\mathrm{pH}\!=\!5.8$, in $10$~mM NaCl solution, at $T\!=\!298$~K, are given in Table~\ref{tab1}. Other chemical reagents, such as sodium chloride, hydrochloric acid, and buffers (Sigma Aldrich) were used without additional purification. Ultrapure water was obtained using the Milli-Q Elix \& Simplicity 185 purification system (Millipore).  Silica-coated AT-cut quartz crystals with a fundamental shear oscillation frequency of $5$~MHz (QSense, Biolin Scientific, Sweden) were used in QCM experiments. Before conducting measurements, the sensors were cleaned in a diluted piranha solution ($1$:$1$:$1$ ratio of deionized water, hydrogen peroxide, and sulphuric acid) for $1.5$ minutes, thoroughly rinsed with ultrapure water, and then immersed in water at $80^\circ$~C for $30$ minutes. Afterwards, the sensors were dried in a stream of pure nitrogen and used immediately after cleaning.  

Diffusion coefficient of particles was determined by dynamic light scattering (DLS) using the Zetasizer Nano ZS instrument (Malvern). The hydrodynamic diameter was calculated using the Stokes-Einstein relationship. The electrophoretic mobility of particles was measured using the Laser Doppler velocimetry (LDV), which employed the same apparatus. Using the electrophoretic mobility data zeta potential of particles was calculated using the Smoluchowski formula.  Zeta potential of silica  was determined via streaming potential measurements, performed according to the procedure described in \cite{zbig13} applying the Smoluchowski formula corrected for the surface conductivity. QCM measurements followed the standard procedure described in \cite{zbig20,jpcc26} using the
QCM instrument (Q-Sense, Biolin Scientific, Sweden). First, a stable baseline in a pure electrolyte of a fixed concentration (either $1$~mM or $10$~mM of NaCl, and $\mathrm{pH}\!=\!5.8$) was established in the QCM-D cell at a constant flow rate (typically $2.5\! \times \! 10^{-3}$ cm$^3$ s$^{-1}$). Then, a particle suspension  of appropriate concentration was flushed through the cell until a stable signal was recorded. Then  the desorption step was initiated, during which a pure electrolyte solution of the same pH and ionic strength was flushed through the cell. Topography of QCM  sensors was determined by atomic force microscopy (AFM) imaging under ambient conditions in semi-contact mode using the NT-MDT Solver BIO device with the SMENA SFC050L scanning head. The root-mean-square (rms) of the sensors was $0.90\pm0.1$~nm; the roughness correlation length and wavelength were $70\pm10$~nm and $120\pm10$~nm, respectively. The average surface height  within  the region  between adjacent peaks  was $0.4\pm0.1$~nm. 
\begin{table*}[tbh]
\begin{tabular}{|c|c|c|c|}
\hline
\makecell{Particle \\ type} &  \makecell{Diffusivity \\ $D_0$ (cm$^2$~s$^{-1}$)} & \makecell{Diameter \\ $2a$ (nm)}  & \makecell{Zeta potential \\ $\zeta$ (mV)}    \\
%\hhline{|=|=|=|=|}
\hline
A32     & $7.5\times10^-7$      &  $32\pm4$     &   $71\pm2$  \\
A71     & $6.9\times10^{-8}$    &  $71\pm5$     &   $74\pm2$  \\
A90     & $5.4\times10^{-8}$    &  $90\pm5$     &   $76\pm2$  \\
A140    & $3.5\times10^{-8}$    &  $140\pm10$   &   $79\pm3$  \\
A360    & $1.3\times10^{-8}$    &  $360\pm30$   &   $97\pm5$  \\
A510    & $9.9\times10^{-9}$    &  $510\pm30$   &   $120\pm3$ \\
\hline
\end{tabular}
\caption{Physicochemical parameters of the polystyrene (PS) nanoparticles used in this work. \label{tab1}}
\end{table*}

Particle surface coverage was determined using the AFM method as previously described in Ref.~\cite{zbig20,jpcc26}. Accordingly, a QCM run was stopped after completing the desorption step, the sensors was removed from the suspension, carefully dried under a controlled humidity and imaged under ambient conditions. The particle surface density, i.e., the number of particles per a unit area (typically one square micrometer) denoted $\widetilde n$, was determined by a direct counting of over a few equal sized areas, randomly chosen over the sensor with the total number of particles about $1000$. All  experiments were carried out  at the temperature of $298$~K, in $10$~mM NaCl solutions of $\mathrm{pH}\!=\!5.8$. 
 
Particle deposition kinetics in QCM cell was also theoretically evaluated in terms of the hybrid convective-diffusion equation and Random Sequential Adsorption (RSA) approach (see \cite{zbig17,zbig00}) using the AFM derived real particle coverage as a control.

Several QCM runs were performed for each particle type in Table~\ref{tab1} using different suspension concentrations, and flow rates. The frequency, $\Delta f$, and the dissipation factor, $\Delta {\mathcal D}$, shifts were recorded at various oscillations overtones ($n\!=\!1,3,5,\ldots$) as a function of time. After the signals attained  stationary values, desorption runs were initiated, in which a pure electrolyte solution of the same pH, ionic strength, and flow rate was flushed through the cell. The frequency shifts were expressed in the overtone-normalized form $-\Delta f/n$ , and the dissipation shifts were converted to the half-bandwidth shifts as $\Delta \Gamma/n =\frac{1}{2} f_0\Delta {\mathcal D}$, where $f_0\!=\!5$~MHz is the fundamental frequency.  After completing each deposition/desorption run, the real particle coverage was determined by AFM. This enabled to confirm the precision of RSA modeling, which furnished the particle mass coverage vs. the adsorption time dependencies in a quasi-continuous manner. Using the experimental results for the overtone-normalized frequency and half-bandwidth shifts, the corresponding (dimensional) impedance components can be readily calculated using the small-load approximation (e.g., \cite{review21}),
\be
\rIm[\mathcal Z]=\frac{\pi \mathcal Z_q n}{f_0} (-\Delta f/n)\,,\quad \rRe[\mathcal Z]=-\frac{\pi \mathcal Z_q n}{f_0} (\Delta\Gamma/n)\,, \label{small_load}
\ee   
where ${\mathcal Z}_q\!=\!8.8 \times 10^6$~kg m$^{-2}$ s$^{-1}$ the quartz shear-wave impedance. 

Considering that in the short-time limit the areal number density of particles, frequency and half-bandwidth shifts, are all linear functions of time, the time-independent (dimensionless) impedance components $\rRe[{\mathcal Z}]/(\eta a {\widetilde n})$, $\rIm[\mathcal Z]/(\eta a {\widetilde n})$ in the limit of vanishing surface coverage were determined from (\ref{small_load}) using the corresponding slopes of $\widetilde{n}(t)$, $-\Delta f(t)/n$ and $\Delta\Gamma(t)/n$ at $t\!\to\!0$ (for further details see \cite{jpcc26}).

\section{Results and discussion}

As was mentioned above, for low values of the particle surface coverage, $\theta\!=\!{\tilde n}\pi a^2\!\ll\!1$, the mutual hydrodynamic interactions between adsorbates can be neglected, and summing up the individual contributions of adsorbates renders the dimensionless excess shear force $F$ (scaled with $\eta a v_0$) to be equivalent to the dimensionless complex \emph{impedance}, ${\mathcal Z}/(\eta a \tilde n)$, probed by the QCM-D device.

The comparison of the imaginary part of the impedance found from Eq.~(\ref{close}) vs. the numerical FEM results for a prescribed proximity to the resonator $\epsilon\!=\!0.1$ is depicted in Fig.~\ref{fig2} below. 
\begin{figure}[htb!]
\vspace{2mm}
\includegraphics[width=0.55\textwidth]{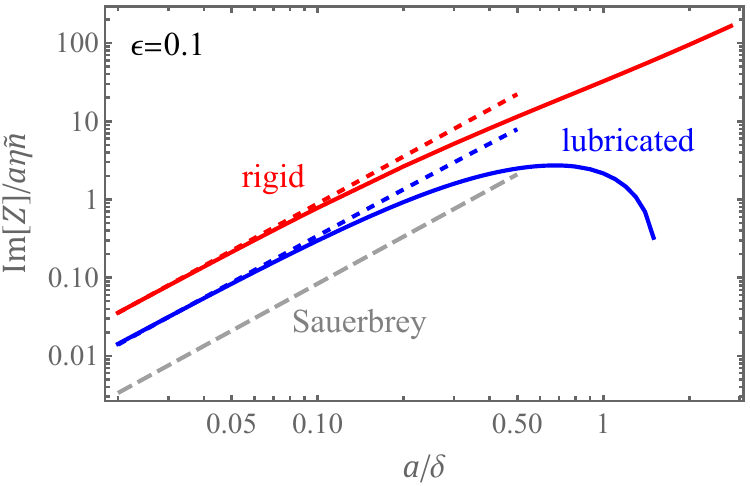} 
\caption{Imaginary part of the scaled impedance, ${\mathcal Z}/(\eta a \tilde{n})$, due to a dilute monolayer of particles ($\theta\!=\!{\tilde n}\pi a^2\!\ll\!1$) with $\rho_s/\rho\!=\!1.05$ and lubricated contact for a fixed minimal separation distance $\epsilon\!=\!0.1$ from the resonator (blue lines) vs. $a/\delta$ (log-log plot). Short-dashed (blue) line stands for the small-$\lambda$ asymptotic approximation (\ref{close}), the solid blue line is the corresponding numerical result; the short-dashed (red) line is the small-$\lambda$ approximate result for the impedance due to a low-coverage monolayer of adsorbates with rigid contact (Eq.~19 in Ref.~\cite{analchem24}) and the solid (red) line is the corresponding numerical result (tabulated in Ref.~\cite{analchem24}). The gray (long-dashed) line stands for the seminal Sauerbrey equation, ${\mathcal Z}_S$, accounting for the particles inertial mass only.  \label{fig2}}
\end{figure}
At this minimal separation, the impedance due to a dilute monolayer of nanoparticles with lubricated contacts is about $2.8$ times smaller than for a layer of rigidly adsorbed particles with stiff contact (red line), and about $3.8$ times larger than that anticipated due to particle inertial mass only (in vacuum) from the seminal Sauerbrey equation,  $\rIm[{\mathcal Z}_S]/(\eta a \tilde n)\!=\! (8\pi/3) (\rho_s/\rho) (a/\delta)^2$ (gray long-dashed line). These results suggest that the QCM signal (i.e., frequency shift) is highly sensitive to the value of $\epsilon$ (please see also Figs.~\ref{fig4}, \ref{fig6} below) and/or the type of the interaction with the resonator surface (e.g., stiff vs. lubricated contact). It will be demonstrated below that for positively charged PS particles adsorbing onto gold/silica sensor, the experimental results follow closely the above theoretical predictions which assumed lubricated contact with the resonator surface.       

To provide further support for the numerical approach, we also compare the numerical results to the prediction of the distant-particle theory \cite{prf23}, that assumes $h\!\gg\! \mathrm{max}(a,\delta)$, while the ratio $a/\delta$ is not constrained, and which uses the reflection method and the representation (\ref{Faltrn}). For the excess shear force due to a stationary particle we previously showed that: 
\be
F_s \approx -\frac{\pi^2 \re^{-2\lambda h} }{\lambda} \left[\frac{3(\re^{2\lambda}-1)}{\pi}+\sum_{l=1}^{\infty}\frac{4(l\!+\!1)I_{l+1/2}(\lambda)}{K_{l+1/2}(\lambda) }\right]\,. \label{Fsfar}
\ee
Here $I_{\nu}(\lambda)$ and $K_{\nu}(\lambda)$ are the modified Bessel functions of the 1st and 2nd kind, respectively. The resistance functions ${\mathcal A}$ and ${\mathcal B}$ have the form \cite{prf23}:
\be
{\mathcal A} \approx 6\pi \re^{\lambda(1-h)}\,,\quad {\mathcal B} \approx -\frac{4 \pi \lambda \re^{\lambda(1-h)}}{1+\lambda}\,, \label{ABfar}
\ee
and the particle velocities read \cite{prf23}:
\be
V \approx \frac{18\pi\,\re^{\lambda(1-h)}}{2\pi(9\!+\!9\lambda+\!\lambda^2)+3\lambda^2\xi}\,,\quad \Omega \approx -\frac{30\pi\lambda\, \re^{\lambda(1-h)}}{20\pi(3\!+\!3\lambda+\!\lambda^2)+3\lambda^2(1\!+\!\lambda)\xi}. \label{VOmfar}
\ee
Substituting (\ref{Fsfar}), (\ref{ABfar})--(\ref{VOmfar}) into Eq.~(\ref{Faltrn}) gives the expression for the excess shear force due to a distant freely suspended particle, which has a particularly compact form for $a\!\ll\delta$: $F\!\approx\! -\lambda^2 \re^{-2\lambda h} (2\pi+\xi)$.

In Fig.~\ref{fig3} we plot the numerical (FEM) results (color solid curves) for the complex linear and angular velocity (their real and imaginary parts) of PS particles ($\rho_s/\rho\!=\!1.05$) plotted vs. the separation distances, starting at $h\!=\!1.002$, for different values of $a/\delta$ varying from $0.1$ to $0.5$. 
\begin{figure}[tbh!]
%\begin{tabular}{cc}
\includegraphics[width=0.9\textwidth]{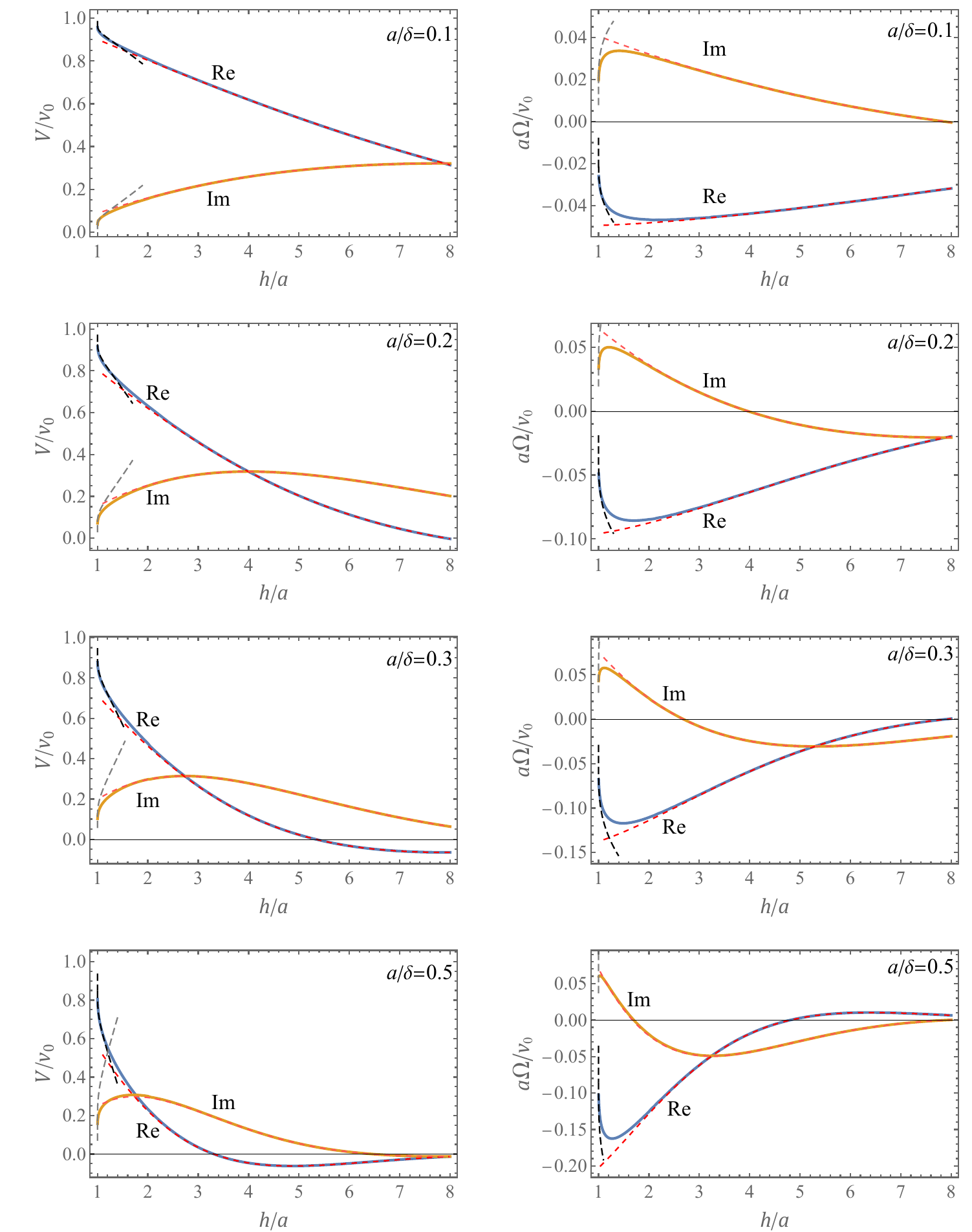} 
\caption{The real and imaginary parts of the linear $V/v_0$ (a) and angular $a\Omega/v_0$ (b) velocities of the PS nanoparticle ($\rho_s/\rho\!=\!1.05$) for different values of $a/\delta$ and lubricated contact vs. the scaled separation distance, $h/a$, from the resonator. Thick solid color lines stand for the numerical results, long-dashed (gray) lines correspond to the small-$\lambda$ asymptotic limit (\ref{U2Omega2}) and short-dashed (red) lines stand for the distant-particle approximation (\ref{VOmfar}). \label{fig3}}
\end{figure}
It can be readily seen that the distant-particle approximation in Eq.~(\ref{VOmfar}) applies already at the distance of $h\sim\!2$ (see the short-dash curves). At $h\!\to\!1$ the near-contact small-$\lambda$ theory in Eq.~(\ref{U2Omega2}) applies (see the long-dash curves), while at contact the velocities $V\!\to\!1$, $\Omega\!\to\!0$ logarithmically slow as predicted by the modified lubrication theory above, solely due to viscous forces (i.e.,  ``viscous adhesion"). Notice that we use the Eq.~(\ref{U2Omega2}) for arbitrarily dense particles, and not the simpler Eq.~(\ref{U1Omega1}) developed for weakly inertial particles, as the latter assumed $a/\delta\!\ll\!0.17$. It can be readily seen that small-$\lambda$ theory provides an accurate asymptotic prediction for moderate $a/\delta\!\lesssim\!0.5$. This is consistent with the findings of Ref.~\cite{prf23} that in the limit of small separations $\epsilon\!\ll\!1$ and $|\lambda|^2 \epsilon \ll 1$, the lubrication equations of the \emph{steady} Stokes flow apply at the leading and sub-leading orders of the asymptotic expansion of the solution in small $\epsilon$, suggesting that it could also hold for $|\lambda|\!=\!{\mathcal O}(1)$. However, not all resistance coefficients ${\mathcal R}_{ik}$ in (\ref{motion}) for moderate $\lambda$ are controlled by the hydrodynamics in the lubrication gap (unpublished) and, therefore, the close agreement between the small-$\lambda$ lubrication theory to the numerical results for $\lambda\!\lesssim\!0.5$ was not readily anticipated.                        

The corresponding impedances, ${\mathcal Z}/(\eta a \tilde{n})$ (its real and imaginary components)  due to a dilute monolayer of nearly neutrally buoyant ($\rho_s/\rho\!=\!1.05$) particles are plotted in Fig.~\ref{fig4} vs. $h/a$ for the same fixed values of $a/\delta$ ranging between $0.1\!-\!0.5$ as in Fig.~\ref{fig3}.  
\begin{figure}[tbh!]
\includegraphics[width=1.0\textwidth]{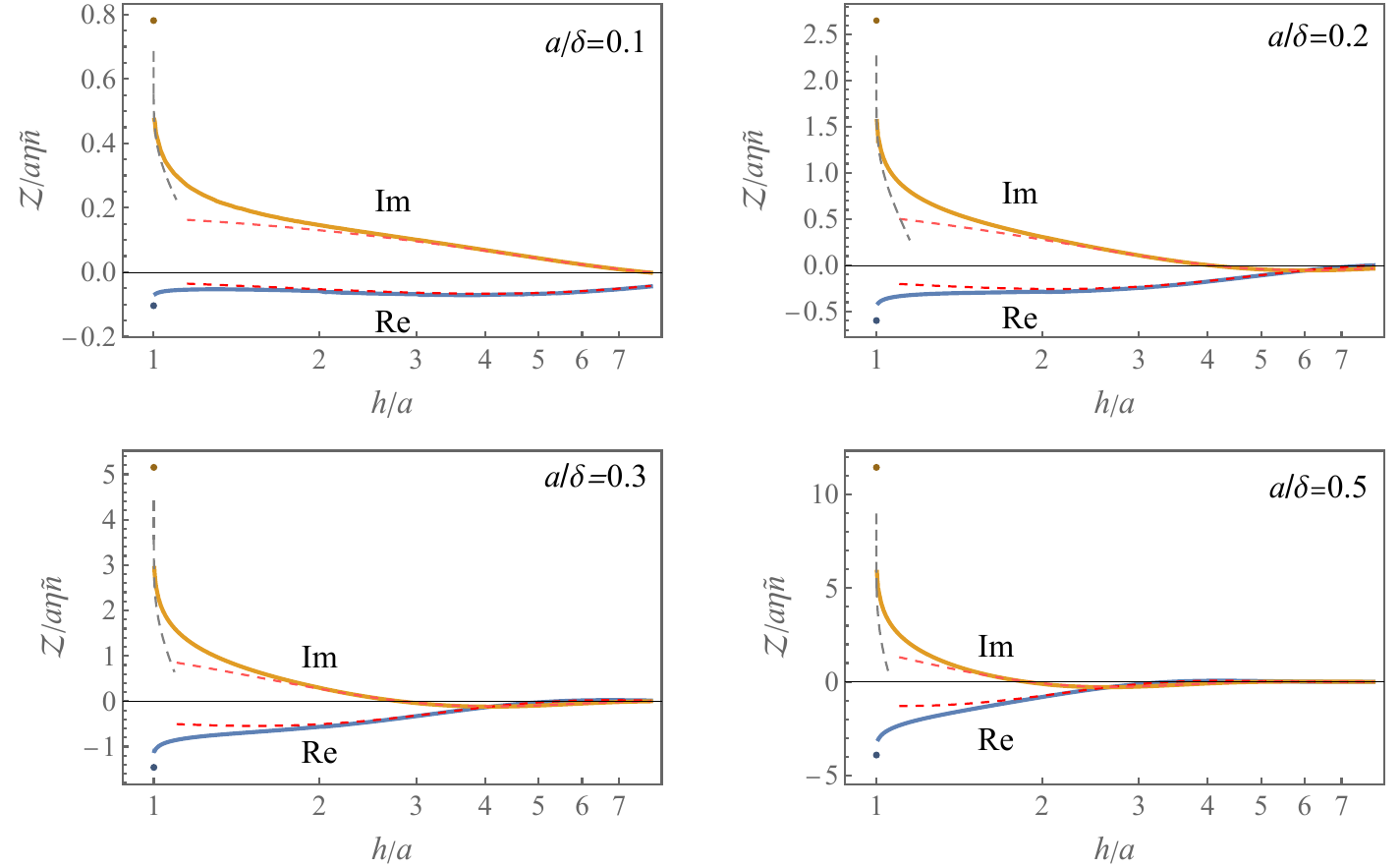} 
\caption{The real and imaginary parts of the scaled impedance ${\mathcal Z}/(\eta a \tilde{n})$ due to a dilute ($\theta\!\ll\!1$) monolayer of PS nanoparticles ($\rho_s/\rho\!=\!1.05$) with lubricated contact vs. $h/a$ (log-linear plot). Thick solid color lines stand for the numerical results, long-dashed (gray) lines correspond to the small-$\lambda$ asymptotic limit for $\rIm[{\mathcal Z}]$ (\ref{close1}) and short-dashed (red) lines stand for the distant-particle approximation (\ref{Faltrn}) using Eqs.~(\ref{Fsfar})--(\ref{VOmfar}) \cite{prf23}; the dots stand for the limiting value of the impedance at $h/a\!\to\!1$ due to rigidly adsorbed particles with stiff contact \cite{analchem24}. \label{fig4}}
\end{figure}
The (red) short dash lines are the predictions of the distant-particle theory, while the long dash curves are the predictions of the near-contact small-$\lambda$ theory in Eq.~(\ref{close1}) for the imaginary part of the impedance. The dots at $h\!=\!1$ denote the limiting values of impedances corresponding to rigidly adsorbed particle. It can be readily seen from Fig.~\ref{fig4}, that the distant-particle approximate theory provides an accurate approximation of the results for $h\!\gtrsim \!2$, while the near-contact asymptotic theory provides an accurate approximation for the impedance near $h\!\approx\!1$, where the FEM computations are challenging to the high spatial resolution of the flow in the thin gap sandwiched between the particle and the oscillating resonator, required for sufficient numerical accuracy.  

Finally, in Figs.~\ref{fig5} we provide the comparison between the theory and the experiments. The symbols represent the scaled impedance due to positively charged amidine polystyrene particles adsorbing onto gold/silica surface. Empty squares stand for the imaginary part and circles for the real component of the impedance, while different colors stand for different particle sizes, ranging from 32~nm (gray) to 510~nm (brown) particles, measured at different oscillation overtones $n\!=\!1\!-\!11$.  
\begin{figure}[tbh!]
\begin{tabular}{cc}
\includegraphics[width=0.48\textwidth]{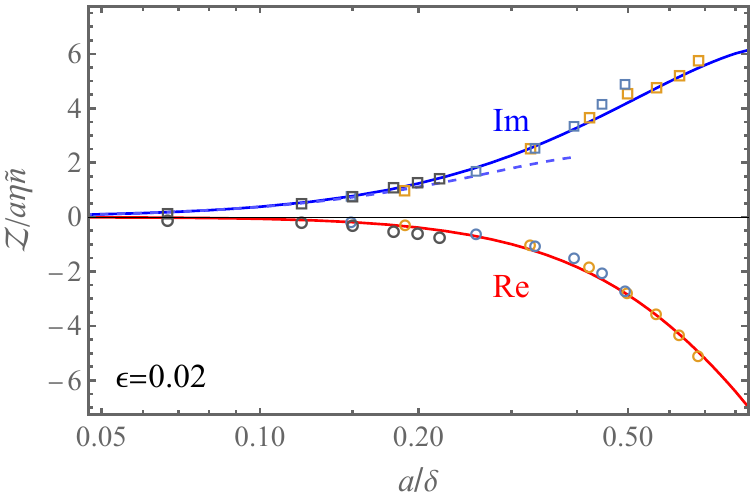} &
\,\includegraphics[width=0.49\textwidth]{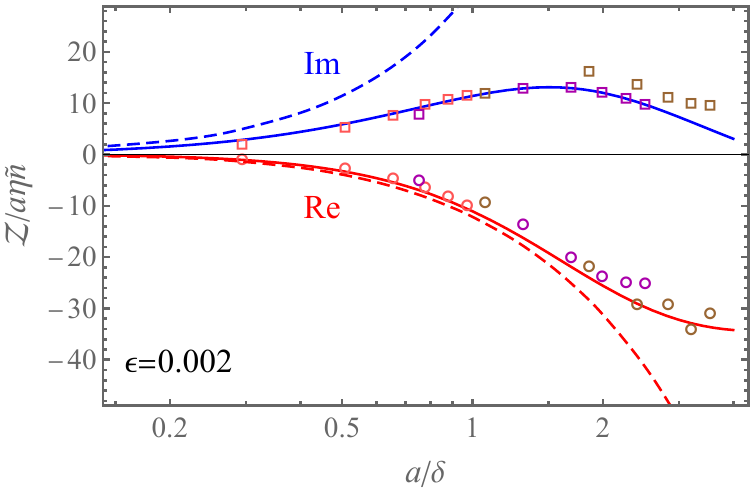} 
\end{tabular}
\caption{The real and imaginary parts of the scaled impedance ${\mathcal Z}/(\eta a \tilde{n})$ due to a dilute ($\theta\!\ll\!1$) monolayer of PS nanoparticles ($\rho_s/\rho\!=\!1.05$) with lubricated contact vs. $a/\delta$ (log-linear plot) for two fixed values of $\epsilon\!=\!0.02$ (left, smaller adsorbates) and $0.002$ (right, larger adsorbates). Thick (color) solid lines stand for the numerical (FEM) results, long-dashed color lines (right plot) to numerical results for rigidly adsorbed particles with stiff contact for comparison. The short-dash line (left figure) correspond to the small-$\lambda$ asymptotic result; the color symbols (circles and squares for real and imaginary parts of $\mathcal Z$, respectively) are the experimental results for positively charged PS nanoparticles in Table~\ref{tab1}: A510 (brown), A360 (magenta), A140 (red); A90(orange), A71 (blue), A32 (gray).  \label{fig5}}
\end{figure}
The solid lines are the results of FEM computations assuming $\epsilon\!=\!0.02$ (left panel) and $\epsilon\!=\!0.002$ (right panel), the long dash curves (right panel) are the numerical results for rigidly adsorbed particles for comparison; the short dash line (left panel) stands for the small-$\lambda$ asymptotic result (\ref{close1}). The respective values of $\epsilon$ for smaller and larger particles roughly correspond to the minimal distance $d\!\approx\!0.7$~nm, which can be estimated as the  sum of the average roughness between peaks equal to $0.4$~nm plus the thickness of hydrated ion layer of $0.3$~nm. For instance for A71 particles we have $d/a\!\approx\!0.019$, while for A510 we find $d/a\!\approx\!0.003$.

The agreement between the experimental data and the theory is very good for both, the real and imaginary components of the impedance without any adjustable parameters, suggesting that PS particles indeed maintain lubricated contact with the gold/silica surface.    

To demonstrate the applicability of the theory, we also performed accurate FEM computations for heavy inertial particles. The particle velocities $V/v_0$ (left panels) and $a\Omega/v_0$ (right panels) in Eqs.~(\ref{U2Omega2}) are depicted (color solid curves) vs. the scaled proximity to the resonator, $h/a$, in Fig.~\ref{fig6} for $\rho_s/\rho\!=\!21.4$, corresponding to platinum (Pt) nanoparticles. 
\begin{figure}[tbh!]
%\begin{tabular}{cc}
\includegraphics[width=0.95\textwidth]{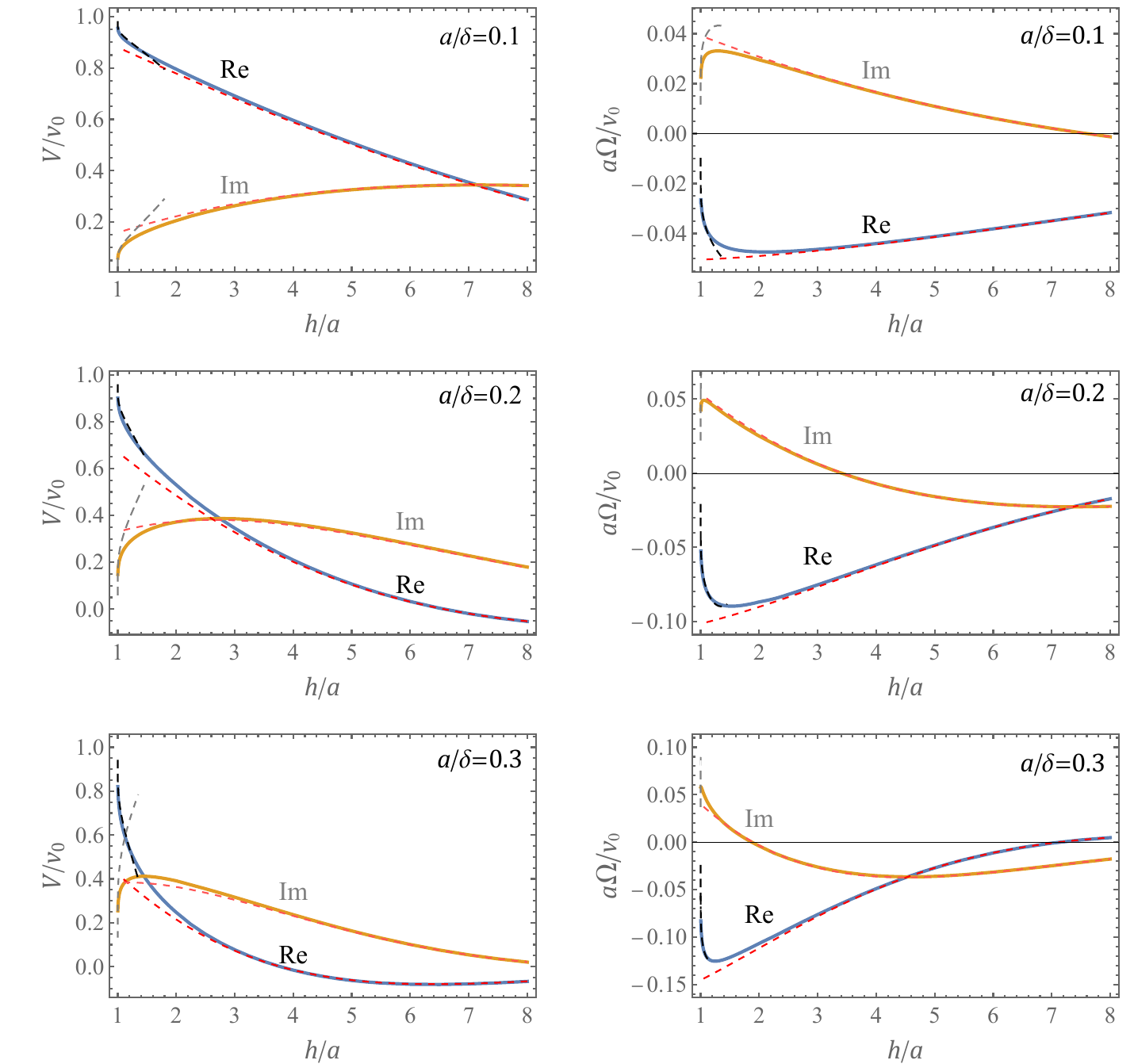} 
%\end{tabular}
\caption{The real and imaginary parts of the linear $V/v_0$ (a) and angular $a\Omega/v_0$ (b) velocities of the Pt nanoparticle ($\rho_s/\rho\!=\!21.4$) and lubricated contact vs. the scaled separation distance $h/a$, from the resonator. Thick solid color lines stand for the numerical results, long-dashed (gray) lines correspond to the small-$\lambda$ asymptotic limit (\ref{U2Omega2}) and short-dashed (red) lines stand for the distant-particle approximation (\ref{VOmfar}). \label{fig6}}
\end{figure}
The short dash curves correspond to the distant-particle approximate theory, while the long dash lines to the near-contact asymptotic predictions. As in the case of nearly neutrally buoyant PS particles, there is an excellent agreement between the distant and near-contact theories and the numerical results. The corresponding real and imaginary components of the scaled impedance are depicted in Fig.~\ref{fig7}, where the color solid lines stand for the numerical results, short dash lines for the predictions of the distant-particle theory, and the long dash curves for the near-contact asymptotic theory; dots stand for the numerical results for rigidly adsorbed particles \cite{analchem24}.  
\begin{figure}[tbh!]
%\begin{tabular}{cc}
\includegraphics[width=0.5\textwidth]{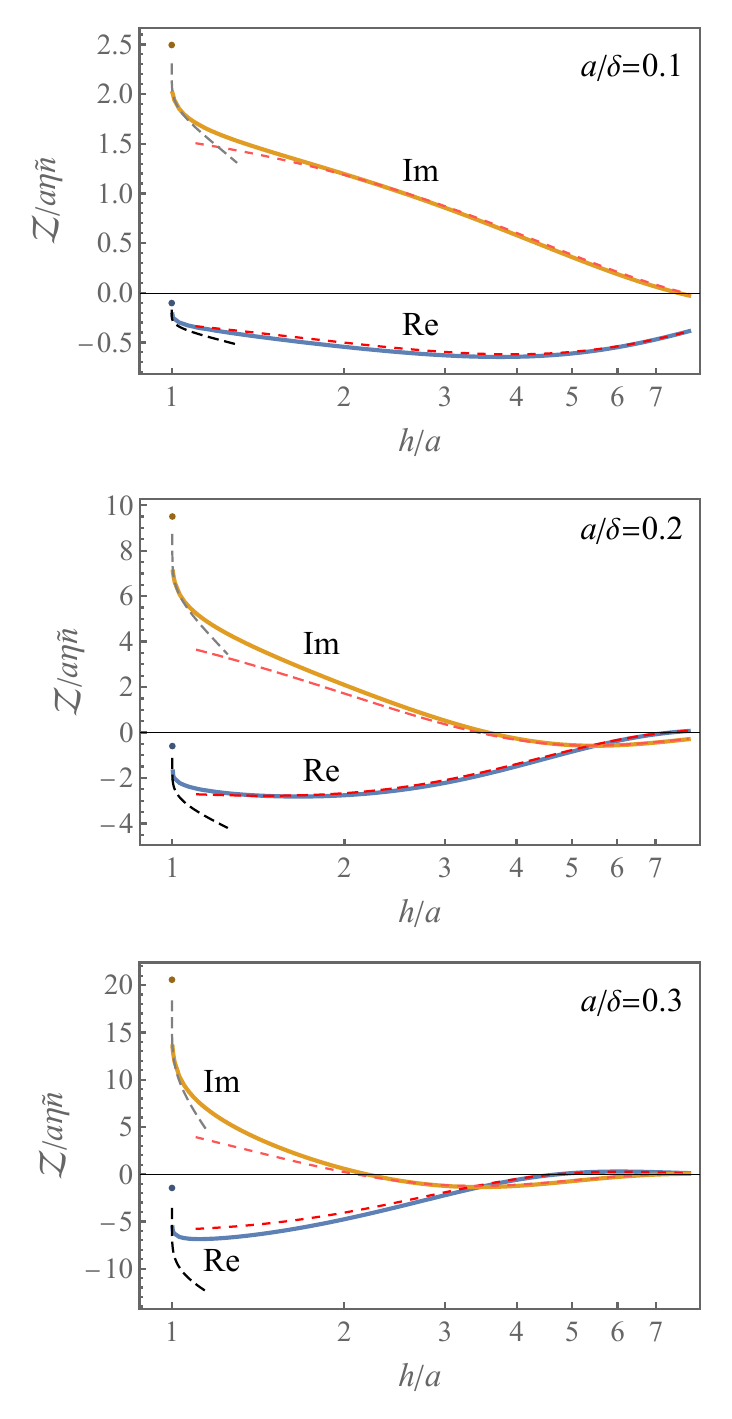} 
%\end{tabular}
\caption{The real and imaginary parts of the scaled impedance ${\mathcal Z}/(\eta a \tilde{n})$ due to a dilute ($\theta\!\ll\!1$) monolayer of Pt nanoparticles ($\rho_s/\rho\!=\!21.4$) with lubricated contact vs. $h/a$ (log-linear plot). Thick (color) solid lines stand for the numerical (FEM) results, long-dashed lines correspond to the small-$\lambda$ asymptotic results for both $\rIm[{\mathcal Z}]$ and $\rRe[{\mathcal Z}]$ and short-dashed (red) curves stand for the distant-particle approximation \cite{prf23}; the dots stand for the limiting value of the impedance at $h/a\!\to\!1$ due to adsorbates with rigid contact \cite{analchem24}. 
\label{fig7}}
\end{figure}
Notice that for heavy enough particles, both the imaginary and the real components of the impedance can be accurately approximated by the asymptotic theory in Eq.~(\ref{close1}) near contact. Notice that opposite to the case of nearly neutrally buoyant PS nanoparticles, the real part of the impedance (i.e., dissipation) due to a layer of heavy inertial Pt nanoparticles with lubricated contact is considerably higher than that due to rigidly adsorbed particles. The detailed comparison of the theoretical predictions to the results of the adsorption experiments involving platinum and golden nanoparticles was reported under a separate cover in Ref.~\cite{jpcc26}.

\section{Conclusions}

In the present paper we theoretically studied a complex impedance due to small (i.e., much smaller than the viscous penetration depth) discrete assuming fluid-mediated (or lubricated) contact with the resonator. The asymptotic limit of small rigidly adsorbed adsorbates (i.e., with a stiff contact) was previously derived in Ref.~\cite{analchem24}. For this purpose we derived a compact representation (\ref{Fcompact}) for the excess shear force exerted on the oscillating resonator in presence of a freely suspended particle, via the Reciprocal Theorem for unsteady Stokes flow, using the problem for a particle adhered rigidly to a resonator as a dual problem. The Eq.~(\ref{Fcompact}) allows to compute excess shear force from knowing the particle linear and angular velocities and quantities corresponding to a dual problem. In the limit of small particle, $a\!\ll\!\delta$, the original unsteady Stokes formulation reduces to a classical problem of steady translation and rotation of a particle suspended in a wall-bounded shear flow, for which the solution is known. Applying with previously derived results for the dual problem in the limit $a\!\ll\!\delta$ \cite{analchem24}, we obtained the closed-form expression for the leading-order excess shear force (its imaginary part) due to a small weakly inertial particle suspended in a close proximity to the resonator, $d\!\ll\!a\!\ll\!\delta$ (\ref{close}). The more general formula accounting for finite particle inertia is also available (\ref{close1}). The resulting asymptotic expressions (i.e., for the complex particle linear and angular velocities and the excess shear force) are in excellent agreement with the accurate FEM numerical computations at small separations $d\!\ll\!a$, while the FEM results were validated vs. the distant-particle asymptotic results corresponding to $h\!\gg\!a$ \cite{prf23}. For heavy inertial particles (such as, e.g., Pt or Au nanoparticles, see \cite{jpcc26}), the asymptotic expression (\ref{close1}) provides an accurate approximation for both, the imaginary and the real part of the excess shear force (see Figs.~\ref{fig7}).   

In the low-surface coverage limit, assuming negligible mutual hydrodynamic interaction of adsorbates, the excess shear force (scaled with $\eta a v_0$) is equivalent to the dimensionless impedance $Z/a\eta\widetilde{n}$ measured by QCM-D. The theoretical (asymptotic and numerical) predictions show an excellent agreement with the experimental measurements for nearly neutrally buoyant polystyrene nanoparticles (of density $\rho_s\!=\!1.05$) of sizes varying between 32~nm to 510~nm, absorbing onto silica sensor at different oscillation overtones $n\!=\!1$--$11$ (see Fig.~\ref{fig5}). Recall that although the excess shear force due to a suspended particle with lubricated contact tends to that of the rigidly adsorbed particle in the limit $d\!\to\!0$ (entirely due to viscosity), it occurs logarithmically slow (see e.g., Fig.~\ref{fig4}), and therefore the resulting (imaginary part of the) impedance reading should be quite sensitive to the value of $d$. The excellent agreement between the experimental impedance measurements and the theory for a wide range of oscillation frequencies and particle sizes, supports unequivocally non-localized adsorption mechanism, i.e. whereas the adsorbate exhibit relative motion (rocking and sliding) with respect to the resonator.            

While many QCM measurements involve small matter, for which $a\!\ll\!\delta$ and the proposed here asymptotic solution applies, however, some other settings involve larger $\mu$m-size objects, e.g., such as bacteria, for which $a\!\gtrsim\!\delta$. In the latter case, $|\lambda|$ is not small and the problem does not reduce to the steady Stokes problem. Moreover, for $a\!\gg\!\delta$, such that $|\lambda|\epsilon\!=\!\mathcal{O}(1)$, the unsteady terms should be retained at the leading order expansion of the lubrication equations, leading to an analog of the Reynolds equation (see, e.g.,\cite{kk91}) governing the pressure distribution in the thin gap of nonuniform thickness separating the two surfaces (unpublished); to the best of our knowledge, this limit has not been considered before and will be studied elsewhere.  

Finally, it is also worth mentioning an experimental technique of AFM force spectroscopy, where the colloidal probe glued to a tip of a cantilever oscillates in a liquid along the direction \emph{normal} to a stationary surface, intended to study the interaction between the probe and the surface (see, e.g., \cite{spectr25b,salez23,salez26}). The quantitative interpretation of the force measurements near contact requires a theory or model for the  hydrodynamic (e.g., added mass and Basset) forces. Apparently, the method developed here could be applied towards AFM force spectroscopy (i.e., at least in the low-frequency limit, $a\!\ll\!\delta$); the corresponding hydrodynamic problem being even simpler than the one considered in the present paper, due to an axial symmetry.

\section*{Acknowledgements}
This work was supported, in part, by the Israel Science Foundation (ISF) via the grant No. 742/25 (A.M.L.) This work was also financially supported by the National Science Centre, Poland, Opus Project, UMO-2026 and partially by the Statutory activity of the J. Haber Institute of Catalysis and Surface Chemistry PAS (M.S. and Z.A.). A.M.L. also acknowledges the support of the David T. Siegel Chair in Fluid Mechanics.

%%%%%%%%%%%%%%%%%%%%%%%%%%%%%%%%%%%%%%%%%%%%%

\end{document}